\documentclass[article]{aa}
\usepackage[varg]{txfonts}
\usepackage{graphicx}
\usepackage{array}
\usepackage{lscape}                                % Tables in landscape
\usepackage{txfonts}
\usepackage{caption}

\usepackage{longtable}
\usepackage{color}
\usepackage{supertabular}
\usepackage[draft]{hyperref}
\usepackage[flushleft]{threeparttable}
\usepackage{multirow}
\usepackage{lscape}

\newcommand{\kms} {km\,s$^{-1}$}

\newcommand{\vsini} {$v$\,sin\,$i$}

\newcommand{\vmacro} {$v_{\rm mac}$}

\newcommand{\Teff} {$T_{\rm eff}$}
\newcommand{\grav} {log\,{\em $g$}}
\newcommand{\gravt} {log\,{\em $g_{\rm true}$}}

\newcommand{\micro} {$\xi_{\rm t}$}
\newcommand{\helio} {Y$_{\rm He}$}
\newcommand{\fastwind} {{\sc fastwind}}
\newcommand{\ioni}[2]{{#1\,\sc{#2}}}

\newcommand{\msol}{$M_{\odot}$}
\begin{document}
%%%%%%%%%%%%%%%%%%%%%%%%%%%%%%%%%%%%%%%%%%%%%%%%%%%%%%%%%%%%%%%%%%%%%%%%%
%
\title{The IACOB project\thanks{Table \ref{tableValues} is available in electronic form at the CDS via \href{http://cdsarc.u-strasbg.fr/viz-bin/cat/J/A+A/638/A157}{http://cdsarc.u-strasbg.fr/viz-bin/cat/J/A+A/638/A157}}
}

\subtitle{VI. On the elusive detection of massive O-type stars close to the ZAMS
%\thanks{}
}

\author{G.~Holgado\inst{1,2,3}, S.~Sim\'on-D\'iaz\inst{2,3}, L.~Haemmerl\'e\inst{4}, D. J.~Lennon\inst{2,3}, R. H.~Barb\'a\inst{5}, M.~Cervi\~no\inst{1}, N.~Castro\inst{6}, \\ A.~Herrero\inst{2,3}, G.~Meynet\inst{4}, J. I.~Arias\inst{5}}

\institute{Centro de Astrobiolog\'ia, ESAC campus, Villanueva de la Ca\~nada, E-28\,692,             Spain
            \and
            Instituto de Astrof\'isica de Canarias, E-38200 La Laguna, Tenerife, Spain.
             \and
             Departamento de Astrof\'isica, Universidad de La Laguna, E-38205 La Laguna, Tenerife, Spain.
             \and
             D\'epartement d'Astronomie, Universit\'e de Gen\`eve, chemin des Maillettes 51, CH-1290 Versoix, Switzerland
             \and
             Departamento de F\'isica y Astronom\'ia, Universidad de La Serena, Avenida Juan Cisternas 1200, La Serena, Chile.
             \and
             Leibniz-Institut f\"ur Astrophysik Potsdam (AIP), An der Sternwarte 16, 14\,482 Potsdam, Germany
            }
		   
\offprints{gholgado@cab.inta-csic.es}

%\date{Date}

\titlerunning{On the elusive detection of massive O-type stars close to the ZAMS}
\authorrunning{Holgado et al.}

%%%%%%%%%%%%%%%%%%%%%%%%%%%%%%%%%%%%%%%%%%%%%%%%%%%%%%%%%%%%%%%%%%%%%%%%%
% \abstract{}{}{}{}{}                           5 {} token are mandatory
%%%%%%%%%%%%%%%%%%%%%%%%%%%%%%%%%%%%%%%%%%%%%%%%%%%%%%%%%%%%%%%%%%%%%%%%%
% 
\abstract
{The apparent lack of massive O-type stars near the zero-age main sequence, or ZAMS (at ages < 2~Myr), is a topic that has been widely discussed in the past 40 years. Different explanations for the elusive detection of these young massive stars have been proposed from the observational and theoretical side, but no firm conclusions have been reached yet.
}
{We reassess this empirical result here, benefiting from the high-quality spectroscopic observations of (more than 400) Galactic O-type stars gathered by the IACOB and OWN surveys. 
}
{We used effective temperatures and surface gravities resulting from a homogeneous semi-automatized {\sc iacob-gbat/fastwind} spectroscopic  analysis to locate our sample of stars in the Kiel and spectroscopic Hertzsprung-Russell (sHR) diagrams. We evaluated the completeness of our magnitude-limited sample of stars as well as potential observational biases affecting the compiled sample using information from the Galactic O star catalog (GOSC). We discuss limitations and possible systematics of our analysis method, and compare our results with other recent studies using smaller samples of Galactic O-type stars. We mainly base our discussion on the distribution of stars in the sHR diagram in order to avoid the use of still uncertain distances to most of the stars in our sample. However, we also performed a more detailed study of the young cluster Trumpler-14 as an illustrative example of how \textit{Gaia} cluster distances can help to construct the associated classical HR diagram.
}
{We find that the apparent lack of massive O-type stars near the ZAMS with initial {\em  \textup{evolutionary}} masses in the range between $\approx$30 and 70~\msol\ still persist even when spectroscopic results from a large non-biased sample of stars are used. We do not find any correlation between the dearth of stars close to the ZAMS and obvious observational biases, limitations of our analysis method, and/or the use of one example spectroscopic HR diagram instead of the classical HR diagram. Finally, by investigating the effect of the efficiency of mass accretion during the formation process of massive stars, we conclude that an adjustment of the mass accretion rate towards lower values than canonically assumed might reconcile the hotter boundary of the empirical distribution of optically detected O-type stars in the spectroscopic HR diagram and the theoretical birthline for stars with masses above $\approx$30~\msol. Last, we also discuss how the presence of a small sample of O2–O3.5 stars found much closer to the ZAMS than the main distribution of Galactic O-type star might be explained in the context of this scenario when the effect of nonstandard star evolution (e.g. binary interaction, mergers, and/or homogeneous evolution) is taken into account.
}
{}


%
%%%%%%%%%%%%%%%%%%%%%%%%%%%%%%%%%%%%%%%%%%%%%%%%%%%%%%%%%%%%%%%%%%%%%%%%%
\maketitle
%%%%%%%%%%%%%%%%%%%%%%%%%%%%%%%%%%%%%%%%%%%%%%%%%%%%%%%%%%%%%%%%%%%%%%%%%
%
%%%%%%%%%%%%%%%%%%%%%%%%%%%%%%%%%%%%%%%%%%%%%%%%%%%%%%%%%%%%%%%%%%%%%%%%%
%%%%%%%%%%%%%%%%%%%%%%%%%%%%%%%%%%%%%%%%%%%%%%%%%%%%%%%%%%%%%%%%%%%%%%%%%
%%%%%%%%%%%%%%%%%%%%%%%%%%%%%%%%%%%%%%%%%%%%%%%%%%%%%%%%%%%%%%%%%%%%%%%%%
%%%%%%%%%%%%%%%%%%%%%%%%%%%%%%%%%%%%%%%%%%%%%%%%%%%%%%%%%%%%%%%%%%%%%%%%%
%
%%%%%%%%%%%%%%%%%%%%%%%%%%%%%%%%%%%%%%%%%%%%%%%%%%%%%%%%%%%%%%%%%%%%%%%%%
\section{Introduction}\label{section1}

A fundamental phase in stellar evolution is the instant of nuclear ignition of hydrogen in the core of the newly formed star. This point in the Hertzsprung-Russel (HR) diagram is commonly known as the zero-age main sequence (ZAMS) and indicates both the beginning of the main sequence (MS) and the theoretical boundary between the star formation process and its further evolution. Another important concept in this context is the so-called birthline, which represents the path followed by a star in formation along the HR diagram until the accreting material is exhausted or until the associated parental cloud is dissolved and the star becomes observable in the optical. The properties of this birthline critically depend on the mass accretion rate during the formation of the star \citep{Bernasconi1996,Norberg2000, haemmerle2019b}. 

If accretion stops before the star has reached the ZAMS, that is, the Kelvin-Helmholtz (KH) timescale of the proto-star is longer than its accretion timescale \citep[as is the case of intermediate- and low-mass stars;][]{Larson1969,Larson1972,Stahler1983},
further evolution of the proto-stellar object until it reaches the ZAMS coincides with a classical KH contraction at constant mass, and is well described by canonical non-accreting models \citep{Siess1997,Baraffe2009,Baraffe2012,Hosokawa2011,Tognelli2015}. 
In this case, the lower envelope of the birthline is delineated by the observable low- and intermediate-mass young pre-MS stellar objects, including Herbig AeBe stars, and different types of of T~Tauri stars \citep[e.g.,][]{Stahler1988,Palla1990,Palla1992,Norberg2000,Behrend2001,haemmerle2019b}.

If, in contrast, nuclear burning of hydrogen starts before accretion is complete \cite[as is the case of stars above a certain mass, see, e.g.,][]{haemmerle2019b}, the star will already have evolved away from the ZAMS when it emerges from the parental cloud. As a consequence, the birthline is expected to delimit a lower envelope (hotter stars) of massive MS stars that are observable in the optical. In this case, when accretion stops and the star has reached its maximum mass, it leaves the birthline to follow the canonical mass evolution toward lower effective temperatures. How close the birthline of massive stars is to the ZAMS depends on the considered accretion rate \citep{Bernasconi1996,Vanbeveren1998,Norberg2000,Behrend2001,hosokawa2009,Hosokawa2010,Haemmerle2016, haemmerle2019b}. 

Stars with masses above $\sim$15~\msol\ are characterized by having O spectral types during their MS phase. Because they evolve rapidly \citep[stars with more than $\sim$40~\msol\ leave the MS in less than 4~Myr; see][ although this strongly depends on specific mass-loss rates]{Brott2011,Ekstroem2012}, the presence of O-type stars in a galactic region normally indicates a recent (or still active) star formation event \citep{Herbig1962}. The short nuclear burning timescale characterizing these stars, which is comparable with the timescale required to dissolve the associated parental cloud, was originally proposed to be the reason of the apparent lack of Galactic O-type stars close to the theoretical ZAMS \citep{Garmany1982}. Other related explanations for this gap in the upper left part of the HR diagram have been put forward. One of these is the occurrence of observational biases caused by the lack of stars from really young clusters (< 1\,--\,2~Myr) in the investigated samples \citep{Herrero2007} and/or the fact that young massive stars, being still embedded in their birth cocoon, are heavily reddened and hence easily elude magnitude-limited samples \citep{Yorke1986,Hanson1998}. However, another interesting possibility has not yet been investigated in detail \citep[in spite of former approaches, e.g., ][]{Herrero2007}: that the empirical hot boundary of O-type stars detected in the optical might correspond to a stellar birthline of massive stars associated with a lower accretion rate than canonically assumed \citep[see, e.g.,][]{Vanbeveren1998}.

The elusive detection of massive stars with mid-O spectral types (i.e., with masses in the range $\sim$30\,--\,60~\msol) close to the ZAMS has been a persistent empirical result since the pioneering work on the Milky Way by \citet{Garmany1982}.
This peculiar empirical feature was also shown to be present (although not in a completely conclusive way) in other galaxies of the Local Group in \cite{Massey1993}, \cite{Massey1995a}, and \cite{Massey1995b}. These authors, who investigated the massive star population of the Magellanic Clouds, NGC~6822, M~31 and M~33, also found that in the case of the Magellanic Clouds, their reddening data rendered the suggestion unlikely that such an absence (if real) would be due to the length of time that it takes a massive star to emerge.

Although the above-mentioned studies based the determination of effective temperatures (\Teff) and absolute magnitudes ($M_{\rm v}$) of their large samples on standard (by that time) $M_{\rm v}$, bolometric correction (B.C), \Teff\ calibrations \citep[][]{Morton1969,Conti1975}, and distance modulus given in the literature \citep[e.g.,][]{Humphreys1978}, similar results were later obtained by different authors using medium-size samples of O-type stars that were investigated spectroscopically, both in the Milky Way \citep{Herrero1992, Herrero2007, Repolust2004, Castro2014, Holgado2018} and the Large Magellanic Cloud \citep{Sabin-Sanjulian2017}. 

In this work, we benefit from (1) the observational efforts devoted in the past decade by the Galactic O Star Spectroscopic Survey \citep[GOSSS,][]{MaizApellaniz2011} and the high-resolution spectroscopic surveys IACOB and OWN \citep[last described in][respectively]{Simon-Diaz2015, Barba2017}, and (2) the availability of stellar atmosphere codes that incorporate the most important physical processes in the modeling of O-type stars \citep[such as the code used in this work, \fastwind;][]{Santolaya-Rey1997, Puls2005} to reassess this intriguing and still unresolved empirical result.

The structure of this paper is as follows. The observations and characteristics of the sample are described in Sect.~\ref{section2}. Sections~\ref{sectionMethod} and \ref{sectionRes} briefly present the method and results associated with the quantitative spectroscopic analysis of the likely single and single-line spectroscopic binaries in our sample, respectively. In Section~\ref{sectionDis} we evaluate potential observational biases and analysis limitations that might affect our results and might explain the non-detection of O-type stars close to the ZAMS, and we examine possible physical explanations for the existence of this gap. Concluding remarks and ideas for further investigation are presented in Sect.~\ref{sectionConcl}.

%%-----------------------------------------------------------------------
%%--------------------           GRAPH BAP_CompGOSC       ------------------------
%%-----------------------------------------------------------------------
%
\begin{figure*}[t]
\centering
\includegraphics[width=0.49\textwidth]{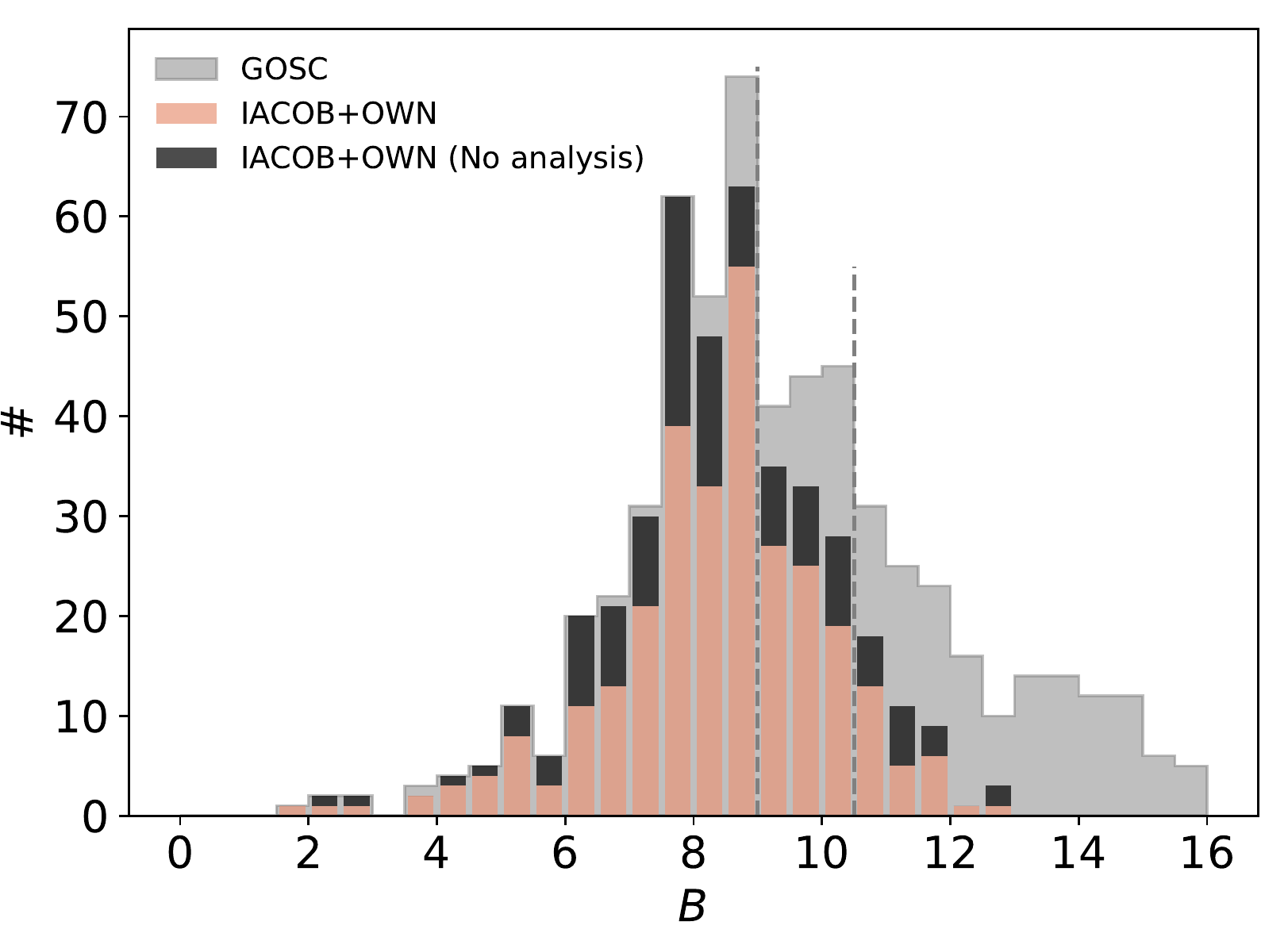}
\includegraphics[width=0.49\textwidth]{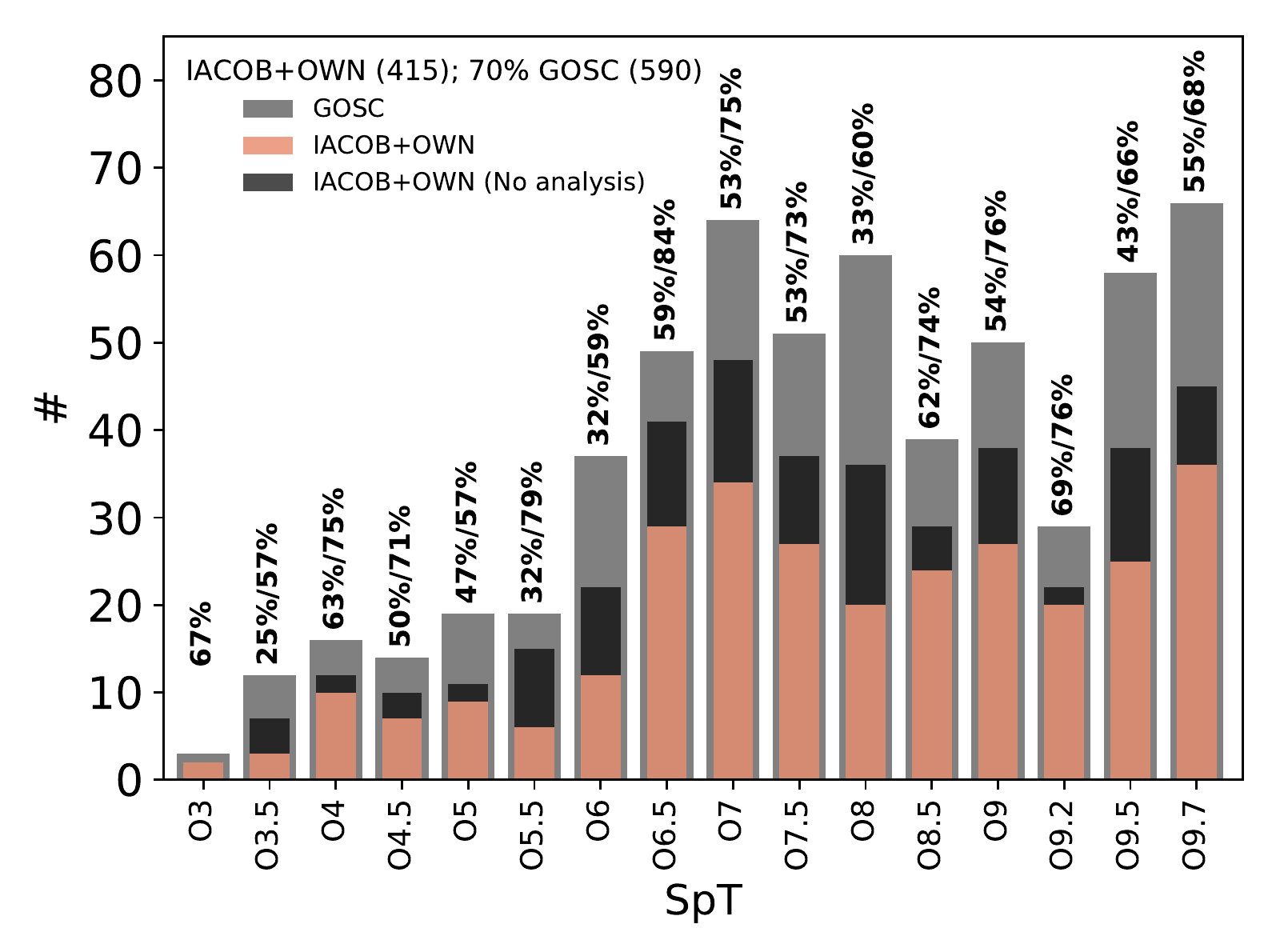}
\caption{Comparison of our working sample of Galactic O-type stars (pink and black) to the complete list of stars of this type included in v4.1 of GOSC (gray). We separate in the two panels the stars for which we have been able to obtain spectroscopic parameters (pink) and those for which we did not perform any quantitative spectroscopic analysis. These last stars are not considered hereinafter (black). \textit{Left} Histogram of stars with respect to the $B$-magnitude. The vertical dashed lines mark the 90 and 80\% completeness limits in our sample using GOSC as reference. \textit{Right} Number of stars per spectral type. Percentages indicate the respective completeness of stars marked in pink and black, in the bottom and top panels, respectively, with respect to GOSC.}
\label{Bap_CompGOSC}
\end{figure*}
%--------------

%%%%%%%%%%%%%%%%%%%%%%%%%%%%%%%%%%%%%%%%%%%%%%%%%%%%%%%%%%%%%%%%%%%%%%%%%
\section{Observations and characteristics of the sample}\label{section2}
%%%%%%%%%%%%%%%%%%%%%%%%%%%%%%%%%%%%%%%%%%%%%%%%%%%%%%%%%%%%%%%%%%%%%%%%%

The observations used for this study came from the two recent spectroscopic surveys IACOB and OWN.  IACOB is a long-term observational project started in 2008 motivated by the compilation and scientific exploitation of a large database of high-resolution multi-epoch spectra of Galactic OB stars. While this project initially concentrated on stars observable from the Roque de los Muchachos observatory (La Palma, Spain), that is, the Northern Hemisphere, in 2012 we established a collaboration with the (also long-term) complementary OWN survey. The latter, started in 2005, pursues a long-term high-resolution monitoring of Southern Galactic O- and WN-type stars. In particular, and for the sake of homogeneity with the spectroscopic observations compiled in the framework of the IACOB project (comprising spectra obtained with the FIES at the NOT2.56m and HERMES at the Mercator1.2m spectrographs), we mainly concentrate on the spectra gathered by the OWN project using the FEROS instrument (attached to MPG/ESO-2.2m).

As described in \cite{Holgado2017, Holgado2018}, the combined efforts of the two surveys have resulted in a high-quality (R$>$25~000,  signal-to-noise ratio, S/N$>$100) database of $\sim$2900 spectra of more than 400 Galactic O-type stars. All these stars are included in version 4.1 of the Galactic O-star catalog \citep[GOSC,][]{MaizApellaniz2013}, from which we also extracted other information of interest for this study, such as spectral classifications, data on photometry and extinction, and some notes on confirmed spectroscopic binarity. We also use GOSC to evaluate the completeness of our sample. In this context, we note that the version of GOSC we used as reference is considered to be complete up to $B$=8. Furthermore, it extends to objects as faint as $B$=16 with a decreasing level of completeness with respect to the expected number of stars per magnitude bin \citep[see Figs. 7 and 4 in][respectively]{MaizApellaniz2013, MaizApellaniz2016}. For example, while the catalog is predicted to be $\sim$95\% complete for stars with a $B$-magnitude in the range 8\,--\,9~mag, this percentage drops to $\sim$50\% in the 9\,--\,10~mag bin (see below and Sect.~\ref{complete}).

As shown in Fig.~\ref{Bap_CompGOSC}, our initial working sample includes 415 Galactic O-type stars covering a range in $B$-magnitude between $\approx$2 and 12.5, and spectral types (SpT) between O3 and O9.7 (left and right panels, respectively). We note, however, that not all stars in this initial sample were used for this study. We concentrate on the 285 stars for which we were able to perform a quantitative spectroscopic analysis (i.e. mainly the likely single stars and the single-line spectroscopic binaries, SB1), and excluded stars that are identified as double-line spectroscopic binaries (SB2) and/or that have strong spectroscopic evidence of being Oe or magnetic stars.

As illustrated in the left panel of Fig.~\ref{Bap_CompGOSC},
the IACOB+OWN sample comprises more than 90\% of the stars with $B$\,$\leq$\,9~mag quoted in version 4.1 of GOSC and includes a large portion ($\sim$70\%) of the stars in this catalog with \textit{B} magnitudes in the range 9\,--\,10.5~mag.
The situation becomes more critical for fainter stars; this is a natural consequence of the observational limitations of the IACOB and OWN surveys. We note that the fainter tail of the \textit{B}-magnitude distribution of stars in GOSC is built using spectra compiled in the framework of the intermediate-resolution ($R$\,=\,2~500) survey GOSSS, which is able to cover a wider range in apparent $B$-magnitude than the IACOB and OWN surveys.

The right panel of Fig.~\ref{Bap_CompGOSC} shows the distribution of our working sample by spectral type, again including a comparison with the stars in GOSC.
The figure also indicates for each spectral type the number and percentage of stars for which we were able to obtain spectroscopic parameters, separating them from those for which we did not perform the quantitative spectroscopic analysis (most of them are SB2). We have high-resolution spectra for $\sim$60\,--\,80\% of the stars in GOSC for each spectral type bin in the histogram, although about 10\,--\,30\% of them belong to the sample of stars for which we did not obtain spectroscopic parameters. Overall, there seems to be no great difference between the distribution of spectral types in the two samples; this indicates that our working sample can be considered a good representation of all Galactic O-type stars comprising version 4.1 of GOSC.

%%%%%%%%%%%%%%%%%%%%%%%%%%%%%%%%%%%%%%%%%%%%%%%%%%%%%%%%%%%%%%%%%%%%%%%%%
\section{Methods}\label{sectionMethod}
%%%%%%%%%%%%%%%%%%%%%%%%%%%%%%%%%%%%%%%%%%%%%%%%%%%%%%%%%%%%%%%%%%%%%%%%%

While we here only discuss the stellar effective temperature (\Teff) and surface gravity (\grav) of our sample, we note that this is only a small subset of parameters that we determined using current methods. The full list of results, including parameters such as projected rotational velocity (\vsini), surface abundances, microturbulence, and the wind parameters that can be determined from the quantitative spectroscopic analysis of the optical spectrum of a O-type stars, will be presented in a forthcoming paper.

We refer to \cite{Holgado2018} for a detailed description of our analysis method, with additional details and examples provided in \citet{Sabin-Sanjulian2014, Sabin-Sanjulian2017} and \citet{Simon-Diaz2017}. In brief, in a first step all the multi-epoch spectra of a given star (we count at least three epochs for 70\% of the stars) are used to provide a first classification in terms of spectroscopic variability (due to, e.g., binarity, pulsations, wind-variability and/or other sources of stellar variability). Then, the spectrum with the best S/N for each star (considering only likely single or single-line spectroscopic binaries, SB1) is used for the quantitative spectroscopic analysis. The latter is performed by means of two semi-automatized tools designed in the framework of the IACOB project, {\sc iacob-broad} \citep{Simon-Diaz2014} and {\sc iacob-gbat} \citep{Simon-Diaz2011}. These two together allow determining the line-broadening and spectroscopic  parameters as well as the associated uncertainties of large samples of O-type stars in a homogeneous, objective, and relatively fast way.

The {\sc iacob-broad} analysis is based on the application of a combined Fourier transform (FT) plus goodness-of-fit (GOF) technique to an isolated line profile. For the sample under study, we used  \ion{O}{iii}$\lambda$5592 as the main diagnostic line, although in a few cases (very fast rotators), we needed to rely on \ion{He}{i} or \ion{He}{ii} lines. {\sc iacob-broad} then provides estimates for the projected rotational velocity (\vsini) and the amount of macroturbulent broadening (\vmacro). 

The remaining spectroscopic parameters, \Teff, \grav, \helio, \micro\ (\textup{microturbulence}), $\beta$ (the exponent of the wind velocity law\footnote{$v(r)=v_{\infty}\left( 1 - R_{*}/r \right)^{\beta} $}), log~$Q$ (wind-strength parameter\footnote{$Q$\,=\,$\dot{M}$~(R v$_{\infty}$)$^{-3/2}$}),  are obtained with {\sc iacob-gbat}. This grid-based automatic tool, which is optimized for the quantitative spectroscopic analysis of O-type stars, performs an optimized $\chi^2$ line-profile fitting of a set of \ion{H}{i} and \ion{He}{i-ii} strategic diagnostic lines \citep[see e.g.,][]{Holgado2018} using synthetic profiles associated with a vast grid of \fastwind\ models \citep{Santolaya-Rey1997,Puls2005,RiveroGonzalez2012}. As indicated above, for this study, we only retained two of the resulting parameters from the {\sc iacob-gbat} analysis: the effective temperature (\Teff), and surface gravity (\grav). The latter was combined with the derived \vsini\ to compute the surface gravity corrected for centrifugal acceleration \citep[\gravt, where $g_{\rm true}=g+g_{\rm cent}$, see][]{Repolust2004}\footnote{Where we estimated the stellar radius using the calibration by \cite{Martins2005}.}.

\begin{figure}[t]
\centering
\includegraphics[width=0.5\textwidth]{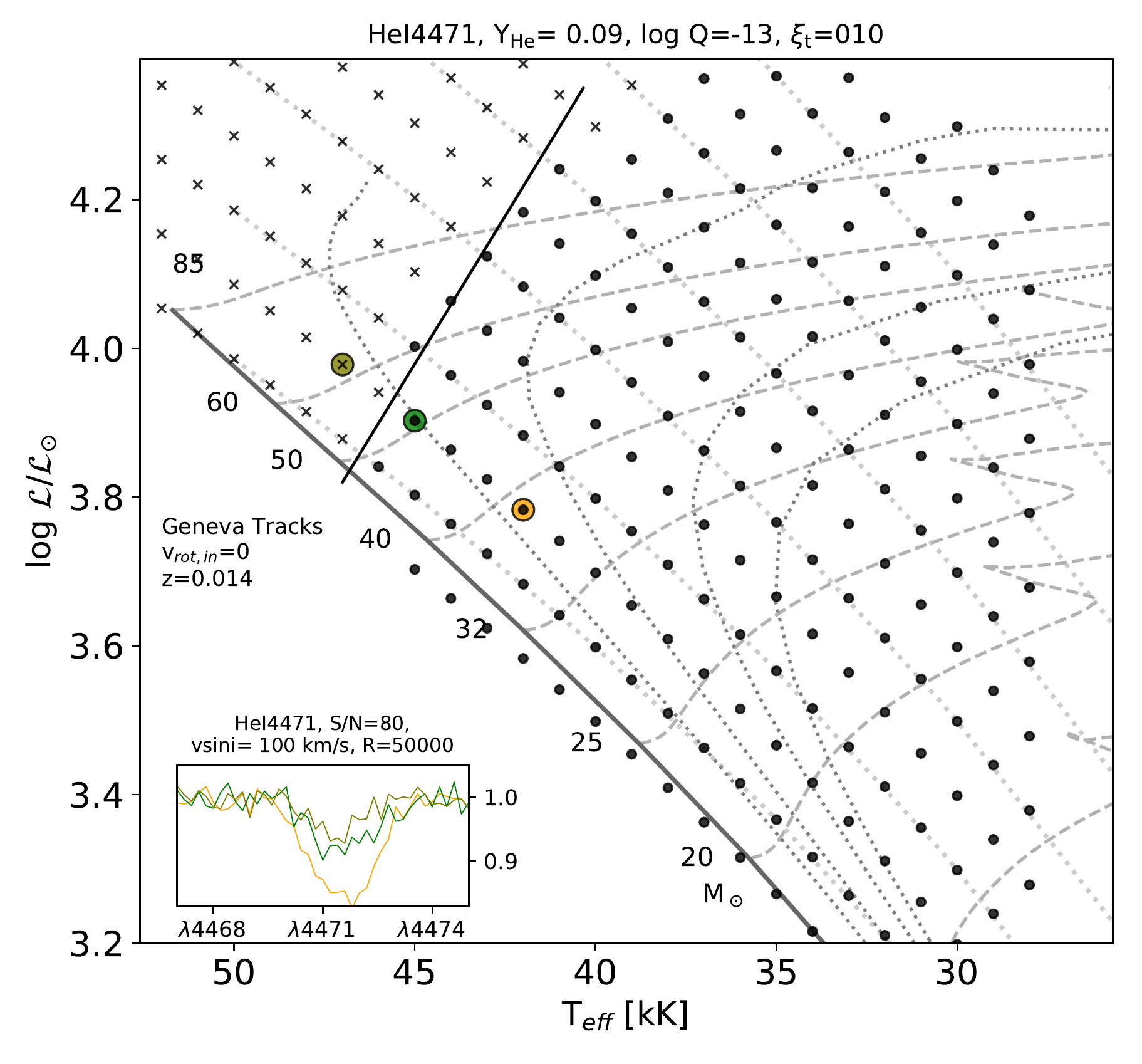}
\caption{Coverage of our grid of \fastwind\ models in the sHRD \citep{Langer2014}. The evolutionary tracks (dashed), location of the ZAMS (solid gray), and isochrones (dotted) for $\tau$=1, 2, 3, and 4 Myr as resulting from the nonrotating solar metallicity models by \cite{Ekstroem2012} and \cite{Georgy2013} are shown for reference. The dotted diagonal lines (following the points) are isocontours of constant gravity. Crosses map models with $EW$(\ion{He}{i}\,$\lambda$4471)\,$<$\,0.15\,\AA. The solid black line separates these models from those for which the \ion{He}{i}\,$\lambda$4471 line has a larger $EW$ (represented by dots). The inset in the bottom left corner depicts three \ion{He}{i}\,$\lambda$4471 line profiles corresponding to the models highlighted in the figure with large colored circles (see text for details of the analysis).}
\label{VGAP_HeI}
%-0.344531 orange
%-0.184998 green
%-0.101175 olive
\end{figure}

Figure~\ref{VGAP_HeI} shows the coverage in the spectroscopic HR diagram \citep[sHRD,][]{Langer2014} of the grid of {\sc fastwind} models that is incorporated in {\sc iacob-gbat.}  This HR-analogous diagram was constructed by combining the \Teff\ and \gravt\ spectroscopic parameters into the $\mathcal{L}$ parameter, which is equivalent to the L/M ratio as $\mathcal{L}$~:=~\Teff $^4 / g_{\rm true} \sim L/M$, establishing a useful diagram to compare observations and evolutionary models regardless of distance and extinction constraints (see Sect.~\ref{sectionRes}). The figure also depicts for reference the ZAMS, the evolutionary tracks, and several isochrones resulting from the nonrotating solar metallicity Geneva models \citep{Ekstroem2012,Georgy2013}. Our \fastwind\ grid properly covers the full MS for stars with masses in the range 20\,--85~\msol.

%%-----------------------------------------------------------------------
%%--------------------           GRAPH GAP HeI       ------------------------
%%-----------------------------------------------------------------------
%
%--------------

%%-----------------------------------------------------------------------
%%--------------------           GRAPH GAP       ------------------------
%%-----------------------------------------------------------------------
%
\begin{figure*}[t]
\centering
\includegraphics[width=0.99\textwidth]{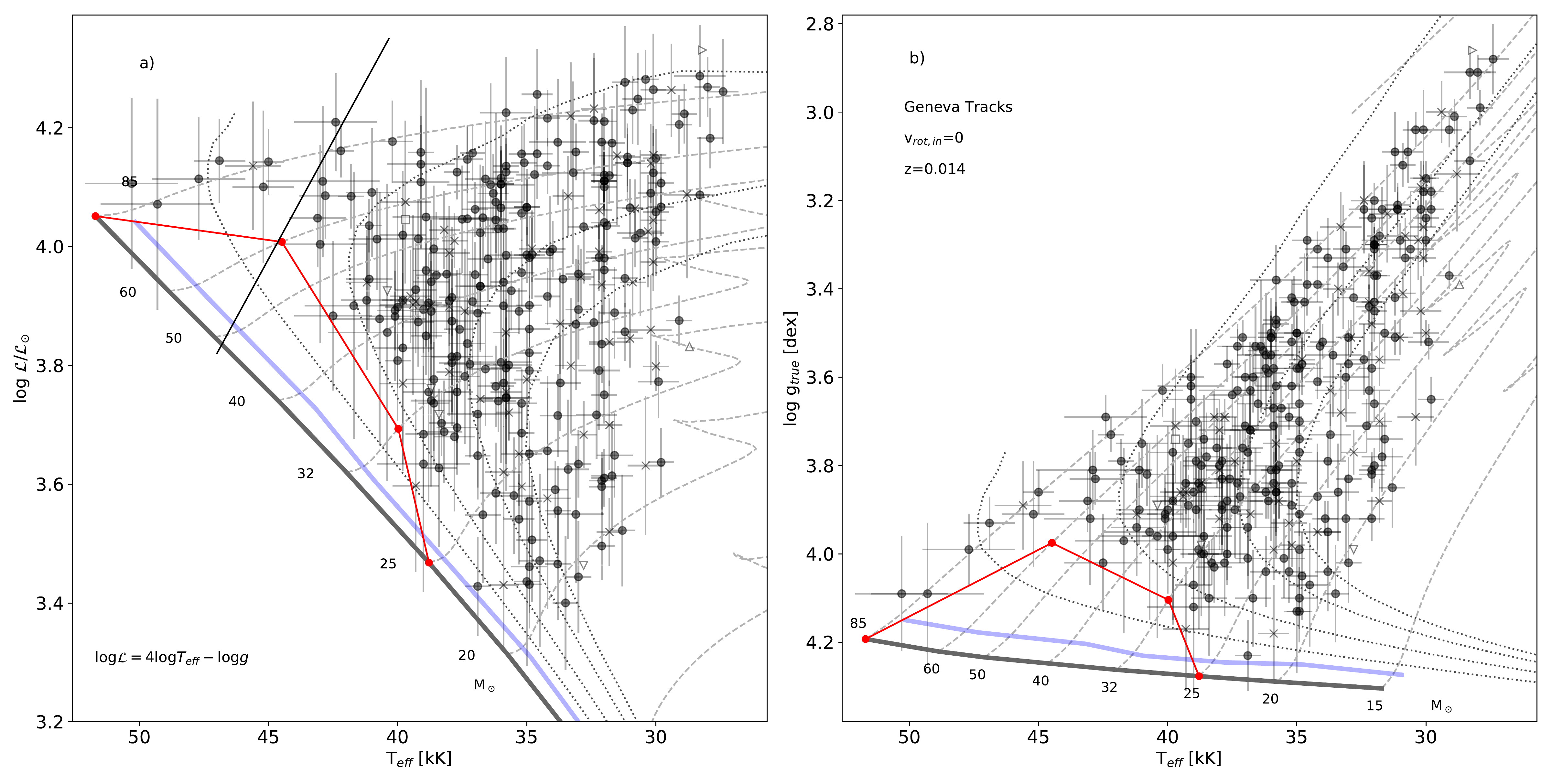}
\caption{Location of 285 likely single and SB1 Galactic O-type stars in the Kiel (\textit{right}) and spectroscopic HR (\textit{left}) diagrams. Crosses indicate stars for which we detected clear or likely signatures of spectroscopic binarity. Open symbols, triangles, and squares are stars for which only an upper or lower limit in any (triangles) or both (squares) of the two parameters used to construct these diagrams could be obtained (\Teff\ and \gravt). Individual uncertainties are included as error bars. Evolutionary tracks and position of the ZAMS (thick solid line) from the nonrotating solar metallicity models by \cite{Ekstroem2012} and \cite{Georgy2013} are included for reference. The thick solid blue line represents the ZAMS for similar models, but with an initial rotation speed of 40\% the critical speed. Isochrones for $\tau$=1, 2, 3, and 4 Myr are also included. The solid diagonal black line separates the region where no \ion{He}{i} lines are available. Last, we also delineate with a red line the region close to the ZAMS where no stars are found.}
\label{VGAP}
\end{figure*}
%--------------

This figure also serves to illustrate a limitation of our analysis strategy.
For very hot stars, the \ioni{He}{i} lines become very weak or disappear (see the bottom left panel in Figure~\ref{VGAP_HeI}). This hampers an accurate determination of \Teff\ based on the \ioni{He}{i} and \ioni{He}{ii} ionization balance. This situation means that the results from {\sc iacob-gbat} in the region of the sHRD that is marked with crosses (and delimited by the diagonal line) are less reliable than for cooler O-type star in our sample. As we show in Sect.~\ref{sectionRes}, the problematic stars are mainly those with spectral types earlier than O4.
An alternative analysis using the \ioni{N}{iv} and \ioni{N}{v} lines \citep[see, e.g.,][]{RiveroGonzalez2012} would be better suited to achieve a more accurate determination of \Teff\ (and hence \grav) for these early O-type stars. 
A complete and detailed HHeN analysis of all O-type stars in the IACOB+OWN sample is planned for a future paper; however, as a sanity check and for the purposes of this paper, we performed a preliminary analysis of the sample of early O-type star using part of the HHeN grid that we are currently computing at the IAC. As a result, we found differences in \Teff\ not larger that 2000 K in several stars, and 1000 K in the rest. This is in part thanks to the availability of the \ion{He}{i}\,$\lambda$5875 line in the {\sc iacob-gbat} analyses, which remains strong enough at higher effective temperatures than the other \ioni{He}{i} lines in the blue region of the spectra.

In this analysis we avoided combining the outcome from our quantitative spectroscopic analysis with data on distances resulting from the parallaxes provided by $Gaia$-DR2 \citep{GaiaCollaboration2018}. While $Gaia$-DR2 data are proving to be of great help to derive more accurate estimates of distances to clusters and associations that include massive stars \citep{Berlanas2019,Drew2018,Drew2019,Davies2019}, the use of individual parallaxes for most of our sample of O-type stars is still problematic because some still limiting systematic errors remain \citep[e.g., calibration limitations for bright stars and undersampling, see][]{Lindegren2018,Arenou2018,Luri2018}. 
In particular, we note that the parallax distribution of our sample is strongly peaked around 0.3--0.4 mas, with a median value of $\sim$0.4 mas. This should be compared with typical parallax uncertainties of 0.03--0.04 mas (see Fig.A.1), and with the position-dependent zero-point uncertainty in the range 0.03--0.08 mas (see the discussion of \citet{Davies2019}). We therefore limit our study to the investigation of the distribution of O-type stars in the spectroscopic HR diagram and defer a discussion of individual parallaxes after publication of $Gaia$-DR3. However, we make use of the stars in the sample belonging to the Trumpler-14 cluster to explore the correspondence between the sHRD and the HRD as an example.

%%%%%%%%%%%%%%%%%%%%%%%%%%%%%%%%%%%%%%%%%%%%%%%%%%%%%%%%%%%%%%%%%%%%%%%%%
\section{Results}\label{sectionRes}
%%%%%%%%%%%%%%%%%%%%%%%%%%%%%%%%%%%%%%%%%%%%%%%%%%%%%%%%%%%%%%%%%%%%%%%%%

Tables~\ref{tableValues}, \ref{tableValues_NoAnalisis}, and \ref{tableValues_NoAnalisis_MagWR} summarize the information about the 415 Galactic O-type stars in our initial sample. Table~\ref{tableValues} includes the 285 stars that we identified as likely single or SB1 for which a quantitative spectroscopic analysis was performed. The other two tables list the targets we identified as SB2 (Table ~\ref{tableValues_NoAnalisis}), or that present features in their spectra that are associated with Oe, Wolf-Rayet, and magnetic stars (Table~\ref{tableValues_NoAnalisis_MagWR}). For the latter two we did not proceed with the spectroscopic analysis, and these objects are accordingly excluded from the discussion presented in Sect.~\ref{sectionDis}. 

In all cases, we quote the spectral classification (this also includes the classification of the secondary component of the SB2 systems whenever available) of each star (as provided in the GOSC) as well as the $B$-magnitude and the E(4405-5495) reddening parameter \citep[extracted from][]{MaizApellaniz2018}. In addition, in Table~\ref{tableValues} we also include estimates and uncertainties for \Teff\ and log~$\mathcal{L}$/$\mathcal{L}_{\odot}$ as resulting from the {\sc iacob-gbat} analysis (see Sect.~\ref{sectionMethod}). In each table, stars are grouped by luminosity class and ordered by spectral type.

The 285 stars for which we were able to obtain the spectroscopic parameters are located in the Kiel (\Teff\ versus \gravt) and spectroscopic HR (\Teff\ versus log$\mathcal{L}$) diagrams in Figure~\ref{VGAP}. Evolutionary tracks and isochrones from the nonrotating models at solar metallicity computed by \citet{Ekstroem2012} and \cite{Georgy2013} are also depicted for reference purposes. Following the discussion presented in Sect.~\ref{sectionMethod}, we draw in both diagrams the diagonal line indicating the boundary between the regions in which strong enough \ion{He}{i} lines are available or are absent. While {\sc iacob-gbat} provides a best-fitting solution for all early O-type stars to the left\footnote{Several cases on the right also have specific characteristics of a helium abundance , \vsini, or S/N that we consider equally limiting to the \Teff\ determination.} of this boundary line, the resulting parameters must be considered with caution (however, see the note in Sect.~\ref{sectionMethod}). All these stars are flagged as ``weak \ioni{He}{i} lines'' in Table~\ref{tableValues}.

The most striking feature in both diagrams is the almost complete absence of stars in the mass range between $\sim$30 and 70~\msol\ to the left of the 2~Myr isochrone (or below, in the Kiel diagram). There is a clear offset that increases with mass between the theoretical ZAMS and the location of the O-type stars in these diagrams. This trend disappears for stars with $\sim$85~\msol, where stars much closer to the ZAMS are found. 

As stated elsewhere \citep[see, e.g.,][]{Repolust2004, Markova2014, Martins2015, Holgado2018}, the Galactic O-type star population is mostly concentrated between the 20 and 85~M$_{\odot}$ evolutionary tracks, and they are basically found to be MS stellar objects (when nonrotating Geneva models are used as reference). However, the expected good coverage of the complete MS domain is challenged by the lack of stars close to the ZAMS. 

Hereinafter, the tracks and ZAMS used for comparison are always those resulting from the set of nonrotating models computed by \cite{Ekstroem2012}. However, in Fig.~\ref{VGAP}, we also show the position of the ZAMS resulting from the Geneva models with $v_{\rm ini}$\,=\,0.4$v_{\rm crit}$ to illustrate that the inclusion of rotation produces a small shift of the ZAMS to cooler temperatures, although this is not large enough to explain the gap between theory and observations. In a future article we will discuss why the set models including an initial rotation of 40\% of the critical speed are not best suited for representing the \vsini\ distribution we obtain for our Galactic sample. This justifies our decision to mainly use the nonrotating models as reference here.

Our result confirms earlier findings by similar spectroscopic studies of  small and intermediate-size samples of Galactic O-type stars \citep[$e.g.$,][]{Herrero1992, Herrero2007, Simon-Diaz2014a, Castro2014, Holgado2018}. It also mimics the results obtained by \citet{Garmany1982}. This empirical result seems to also be present at other metallicities and environments, as is the case for the 30 Doradus region of the Large Magellanic Cloud \citep{Sabin-Sanjulian2017}.

In the next section, we present a thorough assessment of the robustness of the result presented in Fig.~\ref{VGAP} by evaluating (1) the completeness of our magnitude-limited sample of stars, (2) potential observational biases that might affect the compiled sample, and (3) limitations and possible systematics of our analysis method. We then discuss our result in the context of various scenarios proposed so far to explain this elusive detection of O-type stars close to the theoretical ZAMS.

%%%%%%%%%%%%%%%%%%%%%%%%%%%%%%%%%%%%%%%%%%%%%%%%%%%%%%%%%%%%%%%%%%%%%%%%%
\section{Discussion}\label{sectionDis}
%%%%%%%%%%%%%%%%%%%%%%%%%%%%%%%%%%%%%%%%%%%%%%%%%%%%%%%%%%%%%%%%%%%%%%%%%

%-------------------------------------------------------------------------
\subsection{Sample completeness and observational biases}\label{complete}
%-------------------------------------------------------------------------

%%-----------------------------------------------------------------------
%%--------------------           GRAPH Hist_Comp_GOSC_EBV_V       ------------------------
%%-----------------------------------------------------------------------
%
\begin{figure*}[t]
\centering
\includegraphics[width=0.50\textwidth]{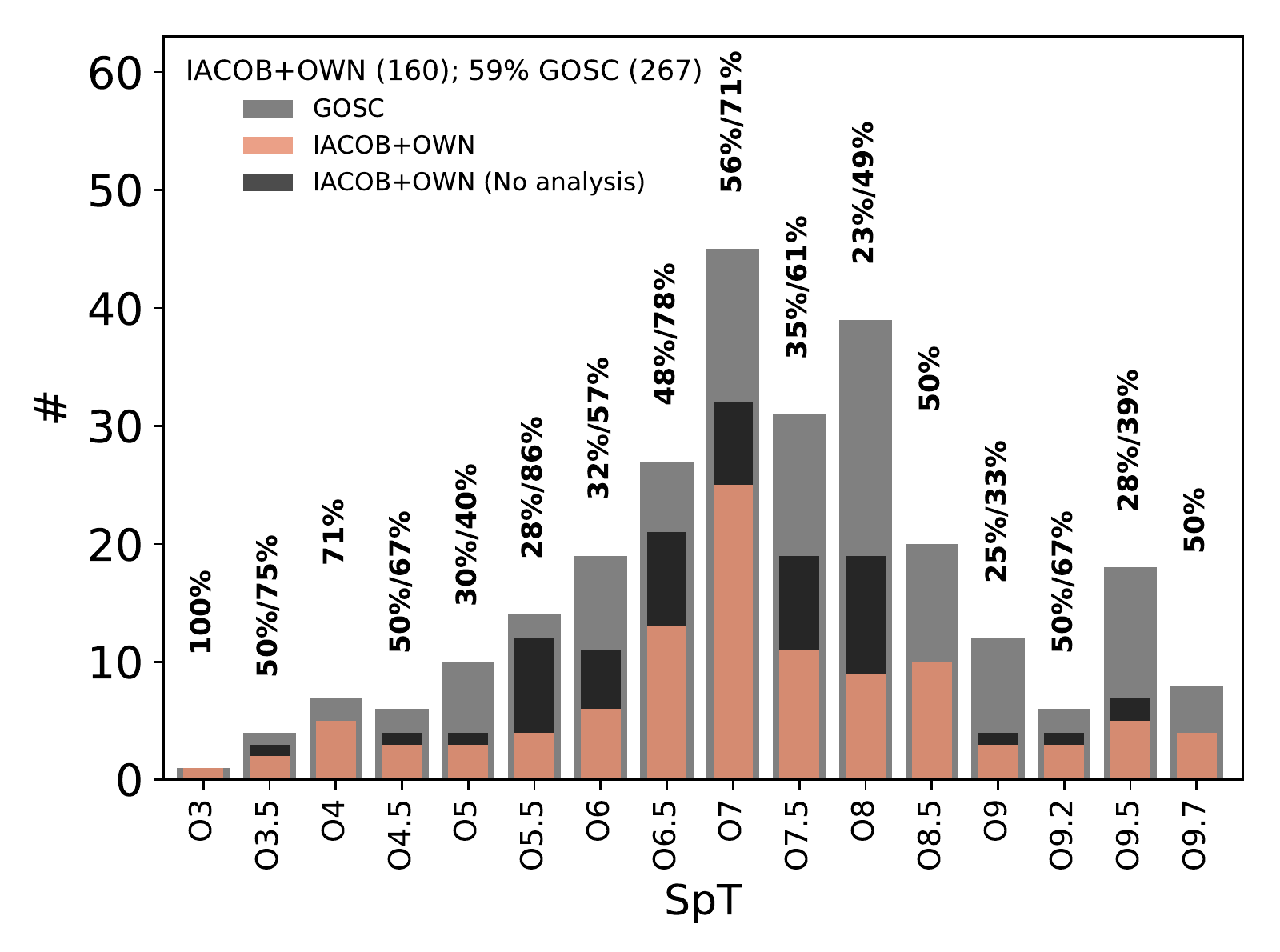}
\includegraphics[width=0.49\textwidth]{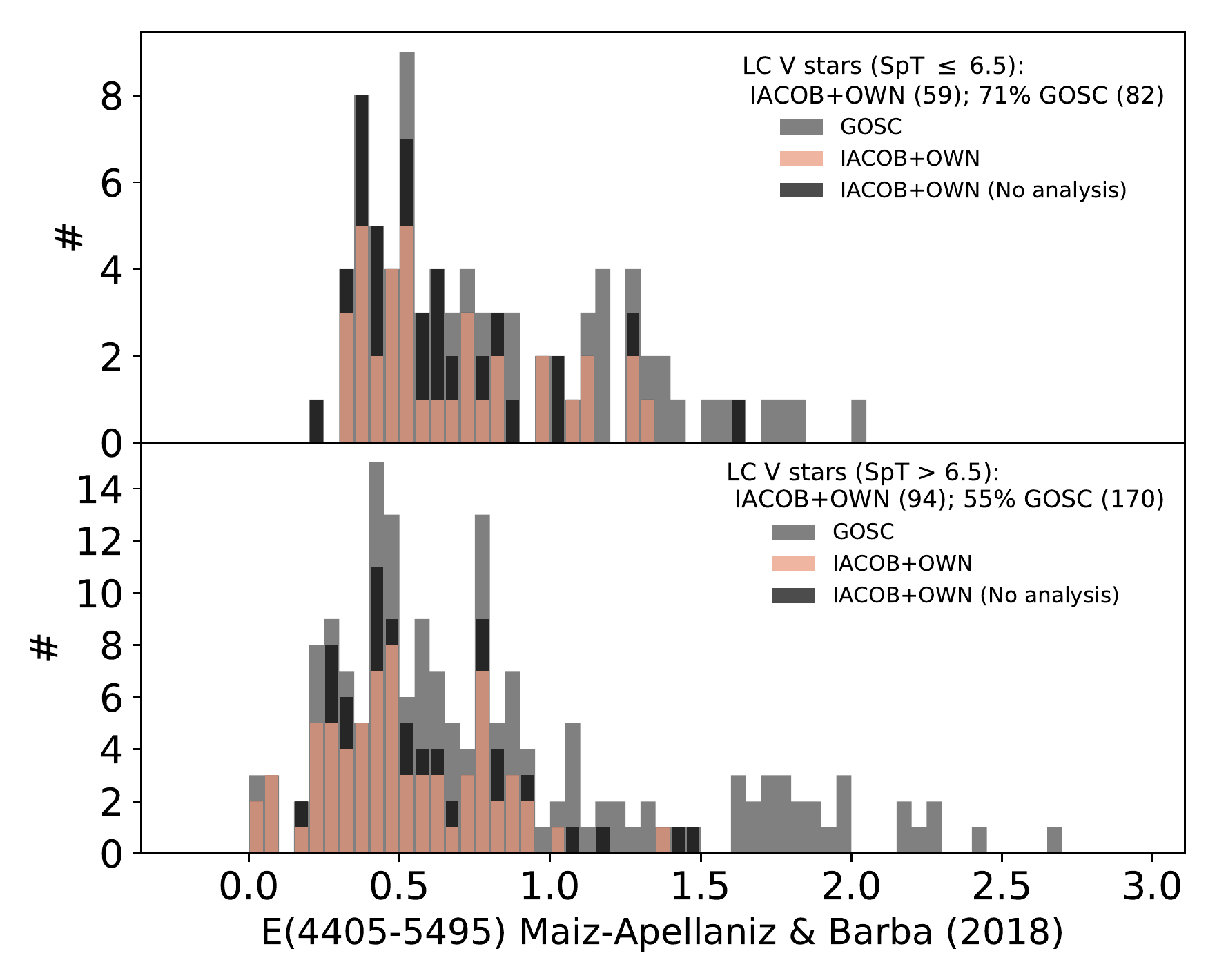}
\caption{Comparison of our working sample of Galactic O dwarfs (pink and black) with respect to those included in GOSC (gray). \textit{Left} Same as Fig.~\ref{Bap_CompGOSC}, but only for luminosity class V stars. \textit{Right} Histograms of the reddening parameter ($E$(4405-5495)$\sim$E($B$-$V$)) separated by ranges in spectral type: earlier (top) and later (bottom) than O6.5.}
\label{Hist_Comp_GOSC_EBV_V}
\end{figure*}
%--------------

The first question we asked ourselves before proceeding with a further interpretation of the results is whether the IACOB+OWN sample may be missing those stars that should be filling the gap. To this aim, we concentrated on the O dwarf population (luminosity class V) and evaluated the completeness of our sample with respect to the GOSC sample (see the left panel in Fig.~\ref{Hist_Comp_GOSC_EBV_V}).

Globally speaking, we assume that $\sim$60\% of the O-type dwarfs are included in GOSC. This percentage is slightly lower than that for the whole sample (see Fig.~\ref{Bap_CompGOSC}), but it still implies that the sample of O dwarfs surveyed by IACOB and OWN is expected to be representative enough for the purposes of our study here. 

The situation further improves when we inspect the critical range of spectral types in more detail. Based on Fig.~\ref{VGAP}, and compared with Fig.~11 in \citet{Holgado2018}, we determine that the missing stars probably have spectral types between $\sim$O6.5 and $\sim$O4.
Because we performed several specific observing runs on which we concentrated on fainter stars than were initially considered in the IACOB and OWN projects, we could increase the number of mid-O-type dwarfs with available high-resolution spectra in these bins by roughly a factor of two, with which we reach $\sim$70\,--\,80\% of the GOSC sample in most cases.
Although the final percentage of stars comprising the analyzed sample (and hence considered in the various sHR diagrams presented in the paper) is somewhat lower than this value (i.e., 30\,--\,50\%, see the bins in pink in the left panel of Fig.~\ref{Hist_Comp_GOSC_EBV_V}), this should not be considered as an indication that our sample is not complete enough for assessing the existence of the gap. This result does not imply that we miss an important fraction of the GOSC stars in these bins, but means that the remaining stars up to the mentioned $\sim$70\,--\,80\% have been identified as SB2 stars.

In the same line of argument, we can now try to answer the question whether based on the approximate volume defined by our sample of dwarfs and our magnitude limit, and assuming that "bare" ZAMS mid-O stars exist within this volume, we might expect to have observed them.
It is straightforward to estimate absolute magnitudes\footnote{We recall that the absolute visual magnitude calibration of O-type stars is observational \citep[see, e.g., ][]{Walborn1972,Walborn1973} and based on stars that have already evolved. It therefore follows that the absolute magnitudes of ZAMS O stars are rather fainter than these calibrations.} for ZAMS stars; we adopted the effective temperature and luminosity at a time step of, for example, 100\,000 years from the evolutionary tracks, and the bolometric corrections of \cite{Martins2006}. Assuming $(B-V)_{0}= -0.28$ mag, as for typical O~V stars, we find that O~V stars between 40 and 60 \msol\ very close to the ZAMS are expected to have an absolute magnitude $M_B=-5.2,-5.7$. Considering a fairly simple average extinction law as a function of galactic position \citep{Amores2004}, this would imply that, for example, with a limiting magnitude of $M_B\sim8,$ we should observe such stars within 2--2.5 kpc, depending on galactic longitude (and assuming low galactic latitude). More specifically, at the distance to Trumpler-14 (parallax of 0.42 mas, see Appendix~\ref{AppTr14} and Sect~\ref{sHRD2HR}), these stars would have apparent magnitudes of $B\sim7.9,7.5$. Even considering a greater extinction, which would produce a reduction of up to two apparent magnitudes, these stars would be marginally included among the stars observed at high resolution, and they would definitely be included in the GOSC sample.

One last question we wish to investigate in this section is whether our sample of stars is affected by the possible observational bias associated with the extinction effect produced by dense material that may be still surrounding the star while the star evolves from the ZAMS. This material is expected to block mostly shorter wavelengths, producing a general reddening trend \citep{Yorke1986,Castro2014}. Therefore, if we are missing O-type dwarfs in GOSC with a high value of the reddening parameter, this may imply that the empty region might be filled by them.

To evaluate this possibility, we present in Fig.~\ref{Hist_Comp_GOSC_EBV_V} two histograms of stars $\text{versus}$ the reddening parameter $E$(4405-5495), as obtained by \citet{MaizApellaniz2018}. We separated the sample into two parts, considering stars with spectral types earlier and later than O6.5, respectively. Although in the sample of late O-type stars we lack the great majority of stars with $E$(4405-5495)\,$>$\,1.0, the expected location of these stars in the spectroscopic HR diagram is below the problematic region (in the sHRD, see Fig.~\ref{VGAP}), and we therefore do not discuss them further. 

For the sample of mid- and early O-type stars, on the one hand, we highlight that of the 19 stars in GOSC with intermediate values of the reddening parameter ($E$(4405-5495)~$\sim 1.0-1.5$), we obtained spectroscopic parameters for 6 of these stars and detected 3 stars as SB2. All of the missing 10 stars are located at a distance \citep[following][]{BailerJones2018} larger than 2.5~kpc that reaches up to 7-8~kpc in some cases. The extinction values of all the missing stars therefore likely come from their large distance.
On the other hand, none of the seven stars with high values of the reddening parameter ($E$(4405-5495)~$\sim 1.5-2.0$) are included in the list of 285 stars for which we have spectroscopic parameters. In addition, distance estimates for these stars are in the range 1.5\,--\,3 kpc; this is not extremely far away. Although this is a small sample, it would be very interesting to obtain spectra and parameters for these stars in the future, with the aim of identifying whether these stars are found closer to the ZAMS than the other stars in our sample. In addition, this study might help to identify potential correlations between high extinction and extreme youth, such that these stars may represent a newborn generation, or at least a less evolved population. In this regard, further exploration of the information available about the local environment (to identify whether these stars are located in dusty and young H~{\sc ii} regions, see also the notes in Sect.~\ref{secVz}) could be informative.

We therefore conclude with relatively good confidence that the existence of a gap is apparently neither due to the fact that our sample of O dwarf stars is small nor to a lack of those stars that should be filling the empty region (at least for likely single and SB1 stars). An in-depth analysis of the stars with very high extinction will shed additional light on the matter.

%-------------------------------------------------------------------------
\subsection{Limitations of the method}\label{methlim}
%-------------------------------------------------------------------------

%%-----------------------------------------------------------------------
%%--------------------           GRAPH GAP Literatura       ------------------------
%%-----------------------------------------------------------------------
%
\begin{figure}[t]
\centering
\includegraphics[width=0.5\textwidth]{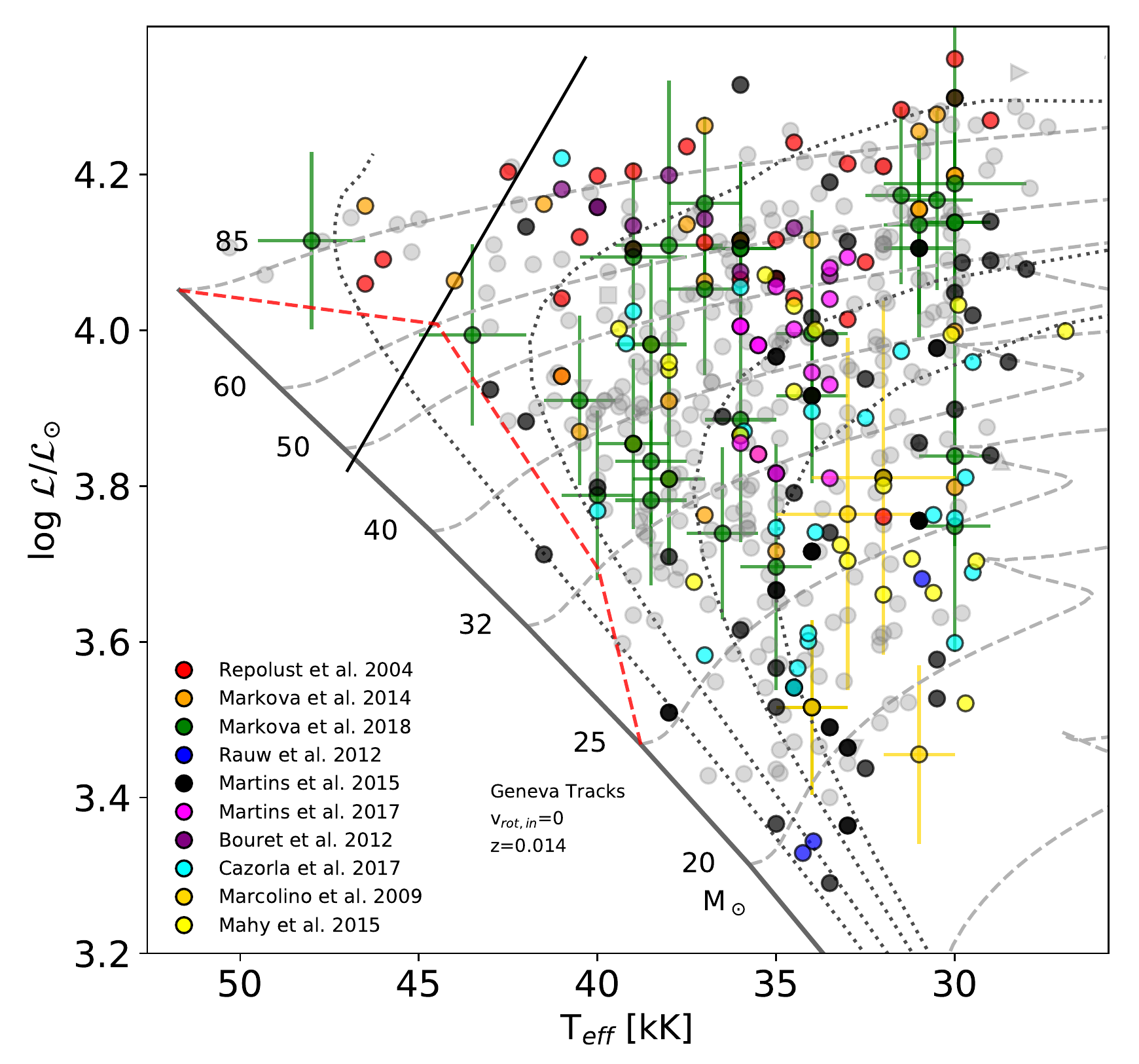}
\caption{Same as Fig.~\ref{VGAP}, but overplotted on our results those obtained by previous spectroscopic studies of small and medium-size samples of Galactic O-type stars found in the literature. Uncertainties are not plotted for cases in which a single value is considered as a standardized error in that particular study.}
\label{VGAP_Lit}
\end{figure}
%--------------

In this section we investigate the possibility that the lack of stars close to the ZAMS in the sHR and Kiel diagrams (Fig.~\ref{VGAP}) is produced by some problem in our method. In particular, we evaluate whether our {\sc iacob-gbat} analysis might erroneously provide gravity estimates below the actual value. This situation may help to fill the gap.

In the past decade, several spectroscopic studies have investigated small and medium-size samples of Galactic O-type stars by means of the modern generation of stellar atmosphere codes and different (but related) analysis strategies. While most of these studies were based on the analysis of the optical spectrum \citep[$e.g.$,][]{Repolust2004,Markova2014,Markova2018,Martins2015,Martins2017,Cazorla2017}, some of them performed a combined optical+UV analysis \citep[][]{Marcolino2009,Bouret2012,Mahy2015}. Figure~\ref{VGAP_Lit} includes the results of all these studies overplotted on our distribution of stars in the sHR diagram. We remark that practically all the stars in the combined sample compiled from the literature are also included in our sample; in addition, we find that our work has allowed us to increase the number of Galactic O-type stars with available spectroscopic parameters by almost a factor 3 (or 5 when we consider the complete IACOB+OWN sample).

Despite the diversity in methods and the use of different stellar atmosphere codes, only one star in the compiled sample is located in the gap region. (However, we note that this star is also included in our analyzed sample and our result differs from the result obtained by \cite{Martins2015}, see Appendix~\ref{AppHD199579}.) This result, together with the relatively good agreement found in \citet{Holgado2018} between the effective temperatures and surface gravities obtained with the \fastwind\ and {\sc cmfgen} codes for a sample of $\sim$100 stars, allows us to stress that the lack of detected O-type stars with ages $\lesssim$1.5~Myr and intermediate masses ($\sim$30\,--\,60~M$_\odot$) is a general outcome of all spectroscopic studies performed to date found in the literature, and it is not necessarily associated with limitations or systematics present in our analysis strategy. Interestingly, our much larger sample has allowed us to better define the gap region by finding a non-negligible number of late O-type stars closer to the ZAMS than were previously found. The results support our argument even more that the gap is not a result of observational biases associated with a magnitude-limited sample because these late O-type stars are $\sim$0.5\,--\,1.5 mag fainter than those that would be occupying the gap region.

%%%%%%%%%%%%%%%%%
We can also wonder whether the solution to this peculiar empirical feature is associated with any missing ingredient in current stellar atmosphere codes. In this line of argument, in a recent study by  \cite{Markova2018}, where the authors reassessed the long-standing mass discrepancy problem \citep{Herrero1992}, it was claimed that part of the reason might be linked to a systematical underestimation of the surface gravities resulting from not accounting from the microturbulent pressure term in the hydrodynamic and quasi-hydrostatic equations when the stellar atmosphere structure was computed. As indicated in \cite{Markova2018}, and to serve as an illustrative example, for a star with \Teff\ = 40 kK, a microturbulence of 15\,--\,20~\kms\ would increase the value of \grav\ by $\sim$0.1–0.15 dex if it were accounted for as a pressure term. 

While this is an interesting hypothesis to be further investigated, it also implies some caveats for the case of the mass discrepancy problem \citep[see notes in][]{Markova2018} and for this paper. For example, if we were to relocate the entire distribution of O-type stars shown in the right panel of Fig.~\ref{VGAP}, some of the late and early O-type stars would be placed below the ZAMS (if the same correction factor were assumed, which may not necessarily be the case). In addition, while this solution may solve both the mass discrepancy problem and the lack of stars close to the ZAMS, it is interesting to note that \cite{Markova2018} found that the mass discrepancy problem is less pronounced precisely in the mass range where the void of O-type stars near the ZAMS appears (between the 32 and 50\msol\ evolutionary tracks).
%%%%%%%%%%%%%%%%%

%%%%%%%%%%%%%%%%%%%%%%%%%%%%%%%%%%%%%%%%%%%%%%%%%
\subsection{More empirical insights}\label{EmpInsi}
%%%%%%%%%%%%%%%%%%%%%%%%%%%%%%%%%%%%%%%%%%%%%%%%%

\subsubsection{Morphological signatures of youth: O~Vz stars}\label{secVz}

The O~Vz phenomenon is a spectroscopic peculiarity defined by a stronger \ioni{He}{ii}\,4686 absorption than in other \ioni{He}{} lines compared to that found in typical class V spectra \citep{Walborn1992,Walborn2009}. This spectroscopic feature was originally proposed to be a clear indication of youth and hence proximity to the ZAMS \citep{Walborn1997}. 
However, as thoroughly discussed in \cite{Sabin-Sanjulian2014}, the situation is more complex, and specific combinations of \Teff, \grav, log~$Q$, and \vsini\ might cause the spectrum of an O dwarf to present the Vz characteristic, in principle independently of age or proximity to the ZAMS. They also note that the Vz characteristic disappears for \Teff\ below 35\,000 K. 

We here followed the quantitative methods defined in \citet{Arias2016} to identify the O\,Vz stars in our sample. We note that the study by \citet{Arias2016} takes into account the results by \cite{Sabin-Sanjulian2014} and provides evidence that these stars are associated with dusty and young H~{\sc ii} regions, for which an independent value of age might be constrained from its morphology.

\begin{figure}[t]
\centering
\includegraphics[width=0.49\textwidth]{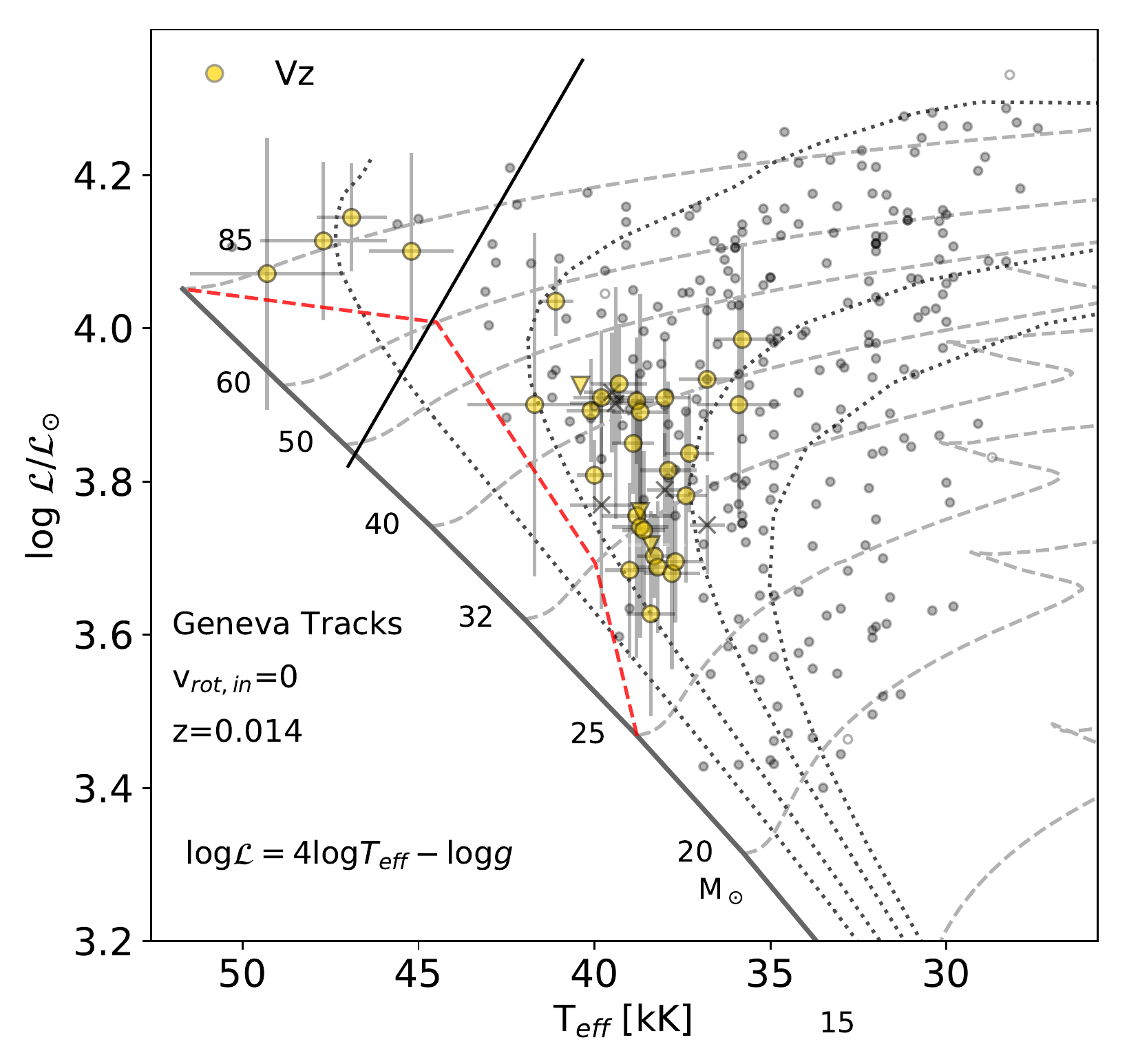}
\caption{Same as Fig.~\ref{VGAP} but highlighting the Vz stars in our sample.}
\label{VGAP_Vz}
\end{figure}

Figure~\ref{VGAP_Vz} shows the distribution of the O~Vz stars in our sample in the sHRD. These stars, highlighted at the top of the complete sample, clearly delineate the young boundary of the distribution of stars ($i.e.$, they are the stars closest to the ZAMS); however, they are still an equivalent of $\sim$0.2~dex away in surface gravity (if we compare them with lines of constant \grav, see Fig.~\ref{VGAP_HeI}) from the theoretical ZAMS defined by the \cite{Ekstroem2012} models. 

%%-----------------------------------------------------------------------
%%--------------------           GRAPH GAP Vz       ------------------------
%%-----------------------------------------------------------------------
%
%--------------

This result once more emphasizes our statement that our sample does not miss O-type stars (detectable in optical wavelengths) that should be closer to the ZAMS due their plausible youth. 

\subsubsection{An extremely young cluster: Trumpler-14}\label{Tr-14}\label{sHRD2HR}

%=============================================================================================================
\begin{table*}[!t]
        \caption{O-type stars in Trumpler-14. For ten stars, high-resolution spectra are available (two of them are SB2 stars), and seven stars are  only listed in GOSC. Columns detail name, spectral class following GOSSS, $B$-magnitude, reddening parameter $E$(4405-5495), \Teff, parameter $\mathcal{L}$, and luminosity, variability notes from this work, and from MONOS \citep{MaizApellaniz2019}. \Teff\ and luminosity errors from {\sc gbat} $\chi^2$ distributions \citep[See][]{Holgado2018}, and limited to 500 K (half of the step in the grid). $\mathcal{L}$ uncertainty from \Teff\ and \gravt\ error propagation.} 
        \label{Tr_14_Stars}
        \centering
        \begin{threeparttable}
            \begin{tabular}{lr@{\hskip 0.05in}lccccccc}
                \hline \hline
        \noalign{\smallskip}
                Name &   SpT &             LC &      $B$ &  E(4405-5495) & $T_{\rm eff}$ &   log $\mathcal{L}$/$\mathcal{L}$$_{\odot}$ & log~$L$   &    Notes       &  Notes   \\
                    &        &                &     [mag]     &         [mag] &     [kK]      &              [dex]                          &  [dex] &  this work   &  MONOS  \\
                \hline
        \noalign{\smallskip}
    \multicolumn{10}{c}{high-resolution spectra available}\\    
    HD\,93129\,AaAb\tnote{a} & O2        &   If*        &  7.8     &  0.514  & 45.6$\pm$1.1 &       4.14$\pm$0.11  & 6.38$\pm$0.03   & SB1  & SB2/3  \\  
    HD\,93129\,B\tnote{a}    & O3.5      &   V((f))z    &  9.0     &     ""  & 47.7$\pm$1.8 &       4.11$\pm$0.10  & 6.43$\pm$0.05   & C    & C      \\  
    HD\,93128               & O3.5      &   V((fc))z   &  9.2     &  0.529   & 49.3$\pm$2.2 &       4.07$\pm$0.18  & 5.78$\pm$0.06   & C    & C      \\
    HDE\,303311             & O6        &   V((f))z    &  9.1     &  0.414   & 40.1$\pm$0.7 &       3.89$\pm$0.07  & 5.16$\pm$0.02   & C    & C      \\
    CPD\,-58\,2611          & O6        &   V((f))z    &  9.9     &  0.562   & 39.8$\pm$0.8 &       3.91$\pm$0.09  & 5.24$\pm$0.02   & C    & SB1?   \\
    HD\,93161\,B\tnote{b}    & O6.5      &   IV((f))    &  8.7     & 0.530   & 37.1$\pm$0.8 &       3.91$\pm$0.13  & 5.74$\pm$0.03   & C    & SB1/2? \\
    HD\,93160\,AB           & O7        &   III((f))   &  8.2     &  0.416   & 36.6$\pm$0.7 &       3.79$\pm$0.10  & 5.68$\pm$0.02   & LPV  & SB1?   \\
    HD\,93161\,A\tnote{b}    & O7.5      &   V          &  8.6     &  ""      &   .          &                   .  & .               & SB2  & SB2    \\  
    Trumpler\,14-9          & O8.5      &   V          & 10.1     &  0.455   & 36.7$\pm$0.7 &       3.55$\pm$0.11  & 5.05$\pm$0.02   & C    & VAR    \\
    HD\,93249\,A            & O9        &   III        &  8.5     &  0.382   &   .          &                   .  & .               & SB2  & SB2    \\
    \hline                                                                   \noalign{\smallskip}   
    \multicolumn{10}{c}{high-resolution spectra not available in IACOB-OWN}\\                                          
    CPD\,-58\,2620          & O7        &   V((f))z    &  9.4     &  0.407    &   .          &                   .  & .    &  .  &  C \\
    ALS\,15204              & O7.5      &   Vz         & 11.5     &  0.787    &   .          &                   .  & .    &  .  &  SB2 \\
    Tyc\,8626-02506-1       & O9        &   V(n)       & 11.5     &  0.591    &   .          &                   .  & .    &  .  &  .  \\
    ALS\,15207              & O9        &   V          & 11.1     &  0.635    &   .          &                   .  & .    &  .  &  . \\
    CPD\,-58\,2625          & O9.2      &   V          & 11.1     &  0.661    &   .          &                   .  & .    &  .  &  VAR? SB2? \\ 
    CPD\,-58\,2627          & O9.5      &   V(n)       & 10.4     &  0.483    &   .          &                   .  & .    &  .  &  . \\
    HDE\,303312             & O9.7      &   IV         & 10.3     &  0.547    &   .          &                   .  & .    &  .  &  Probable SB2  \\
        \hline
        \end{tabular}
                \begin{tablenotes}
                 \item[a \& b] Only one entry for E(4405-5495) in \cite{MaizApellaniz2018}\\
               C: Constant, LPV: Line profile variability, VAR: variability
                 %\item[b] 
                \end{tablenotes}                
     \end{threeparttable}
\end{table*}
%=============================================================================================================

One possible inference from Fig.~\ref{VGAP} is that our data set does not contain clusters that  are young and massive enough to populate the gap. For example, except for the most massive stars in the sample, an isochrone with an apparent age of $\sim$2 Myr is a reasonable fit to the lower envelope of the O stars, particularly around mid-O spectral types. Interestingly, most of the very massive stars that are close to the ZAMS lie within the Carina region, which is one of the youngest and most massive star-forming regions in the Galaxy. 

Carina is a complex region that hosts a number of clusters. Two in particular, Trumpler-14 and Trumpler-16, are thought to be very young, with ages in the range 1--3 Myr \citep{walborncarina,Massey1993,smith2006,Hur2012,damiani}. As discussed in the literature, Trumpler-14 in particular is generally determined to be the younger of the two, based largely on the presence of more-evolved massive stars in Trumpler-16.  \citet{Sana2010} have estimated an age of 0.4 Myr from the study of its pre-MS stars, thereby providing us with an age that is independent of the MS evolutionary tracks we wish to test.
Clearly, the stars in our sample that belong to this extremely young cluster provide powerful diagnostic insight into the gap between models and observations.

However, because the two clusters lie very closely together on the sky, the cluster members might be confused, see, for example, the discussion in \cite{walborncarina} and \citet{smith2006}, and care must therefore be taken to identify bona fide members of these two clusters. In Appendix~\ref{AppTr14} we therefore discuss the Trumpler-14 and Trumpler-16 cluster membership for stars in GOSC, based on \textit{Gaia}-DR2 \citep{GaiaCollaboration2018,Lindegren2018} spatial and dynamical characteristics. We find 17 stars in Trumpler-14 and 20 stars in Trumpler-16 (whereas GOSC lists 14 and 24 members, respectively).

Ultimately, we decided to work solely with the Trumpler-14 sample (see Table~\ref{Tr_14_Stars}), because (a) we had more stars with spectroscopic analysis results (eight versus seven for Trumpler-16), (b) they included most of the very massive stars located in the seemingly separated group in the upper left part of the sHRD, and (c) Trumpler-14 is extremely young.
We note that this list is not intended to be complete, rather it is intended to represent stars that based on their spatial and dynamical properties, are high-probability members of Trumpler-14.

\begin{figure*}[t]
\centering
\includegraphics[width=0.9\textwidth]{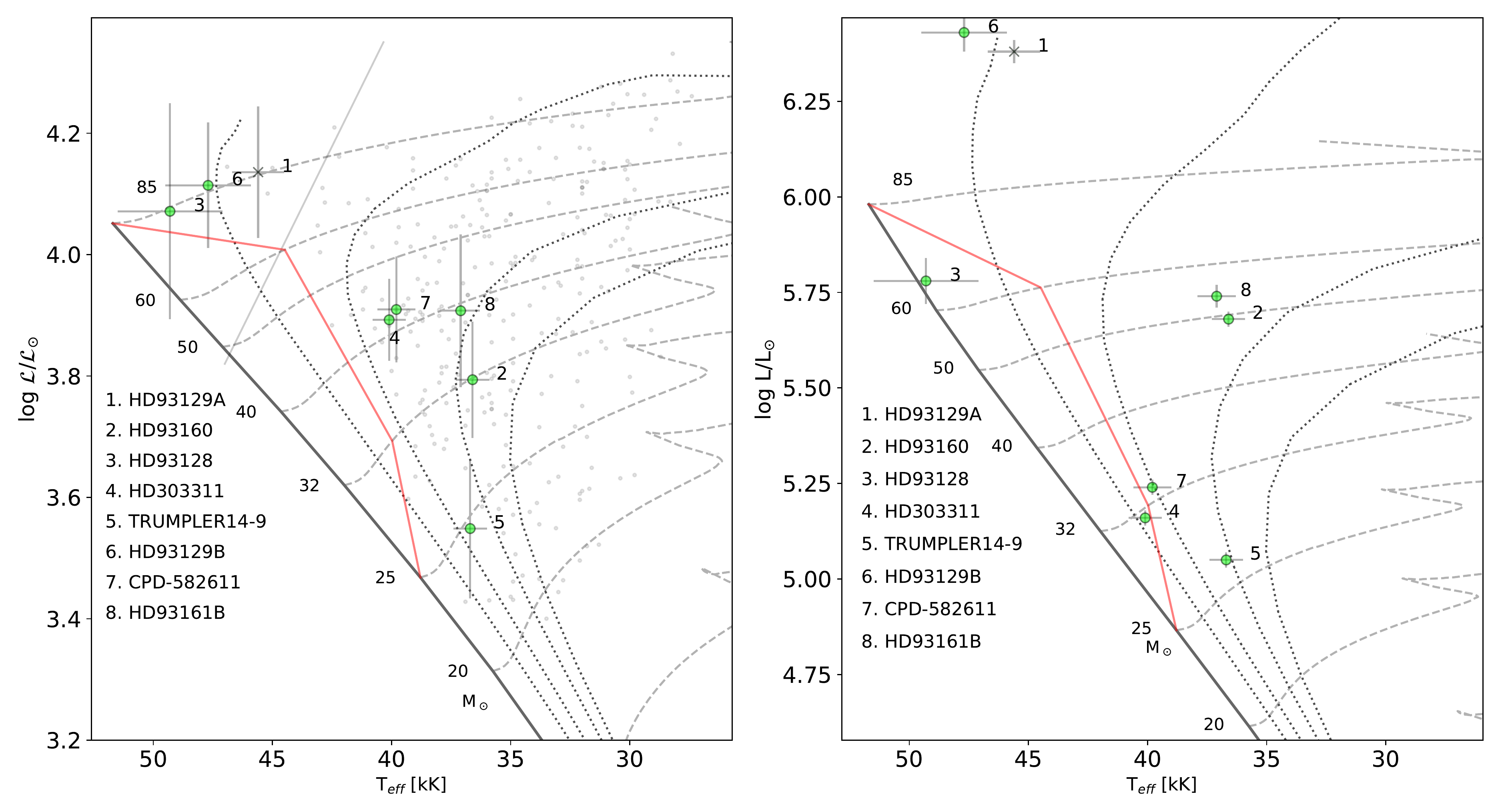}
\caption{Location (and identifier) of the eight Trumpler-14 stars with available spectroscopic parameters in our sample in the sHR (\textit{left}) and original HR (\textit{right}) diagrams. The sHR diagram includes the remaining sample for comparison. Crossess indicate stars for which we have detected clear or likely signatures of spectroscopic binarity. Individual uncertainties are included as error bars. Evolutionary tracks and the position on the ZAMS from the nonrotating solar metallicity models by \cite{Ekstroem2012} and \cite{Georgy2013} are included for reference. Isochrones for $\tau$=1, 2, 3, and 4 Myr are also included. A solid light gray line is drawn to separate the region where no \ion{He}{i} lines are available. Using selected points of the evolutionary tracks in the sHRD to delineate the gap of stars near the ZAMS (see Appendix~\ref{AppRedline}), we are able to transport an analogous region to the HR diagram (red line).}
\label{VGAP_Trumpler}
\end{figure*}

In Fig.~\ref{VGAP_Trumpler} we present the sHRD distribution of Trumpler-14 stars with spectroscopic results with respect to the rest of the sample. For comparison, the tracks provided by the Geneva models \citep{Ekstroem2012} are included. Again, we see the same result as with the complete sample. There are no stars near the ZAMS between the tracks of 32 to 60 \msol. The presence of the moderately large group of very high-mass stars (3 stars with $\sim85$ \msol) so close to the ZAMS indicates a very young age, in agreement with \citet{Sana2011}, but it is noticeable that there are no intermediate-mass stars close to the ZAMS. 

\begin{figure*}[t]
%\centering
\hspace{-2cm}
\includegraphics[width=1.23\textwidth]{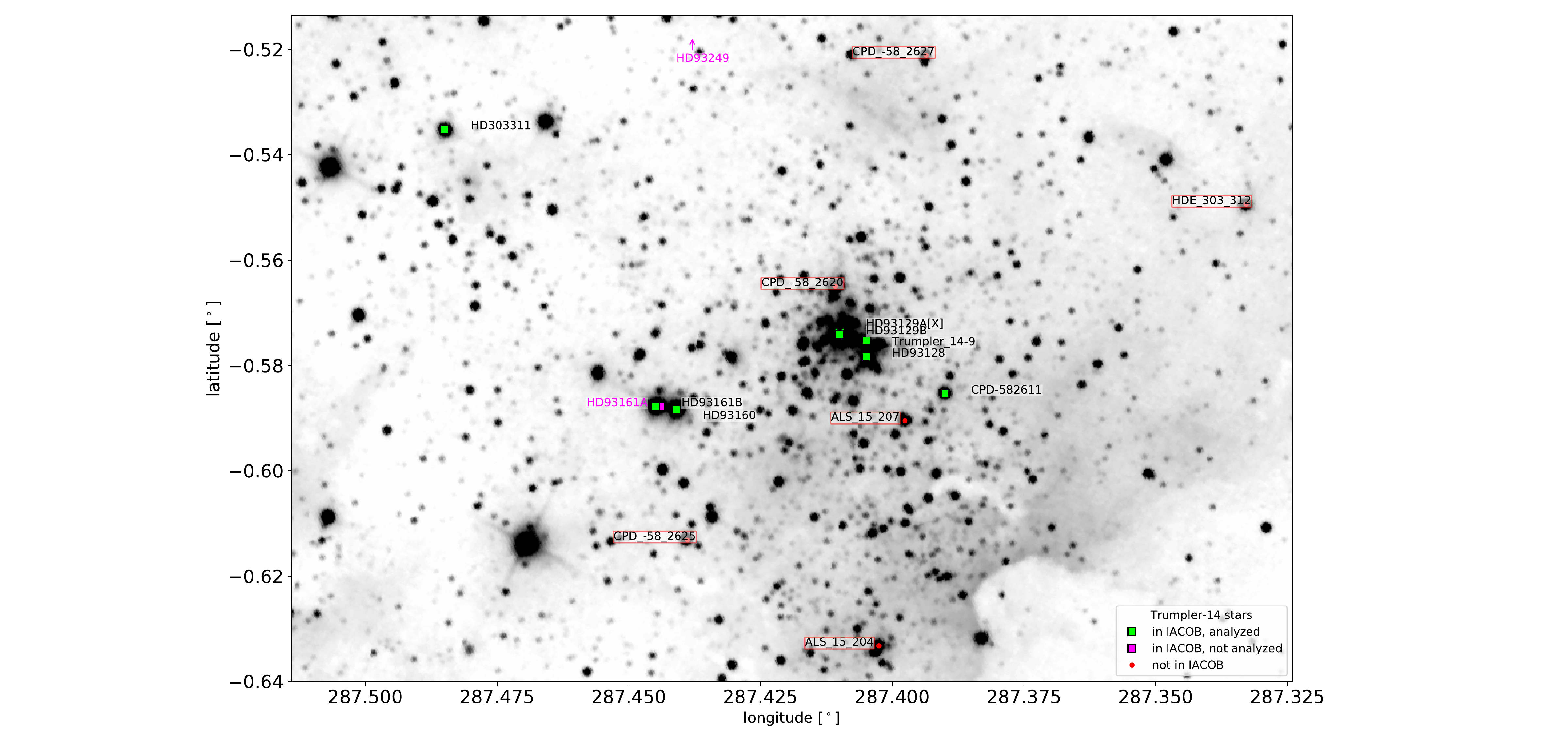}
\caption{Inverse 2MASS J-filter image showing a 10' FoV of Trumpler-14, with the position and name of O-type stars in the sample assigned to the cluster (see Sects.~\ref{secVz} and \ref{Tr-14} for further details). Different colors represent different status: green represents stars with spectroscopic parameters, magenta are SB2 stars for which no quantitative analysis is possible, and red marks stars without high-resolution spectra.
}
\label{Sky_Trumpler_VH}
\end{figure*}

Figure~\ref{Sky_Trumpler_VH} shows an area of $\sim$10' around the center of the Trumpler-14 cluster and depicts GOSC stars that we have considered to belong to Trumpler-14 for which we have IACOB results as well as those for which we do not have high-resolution spectra or spectroscopic results. We have stellar parameters for half of the stars (8 out of 17) that also represent the most cohesive group, with a number in the center of the cluster. For the other 9 stars without parameters, we have a spectral type classification from GOSSS, showing us that only two stars have spectral types O7-7.5 and the rest have spectral types later than O9. These stars are not expected to be able to fill the gap region. Therefore our sample is not biased with respect to GOSC in terms of spectral type, and is representative for the cluster's youth.

%%%%%%%%%%%%%%%%%%%%%%%%%%%%%%%%%%%%%%%%%%%%%%%%%%%%%%%%%%%%%%%%%%%%%%%%%

%%-----------------------------------------------------------------------
%%--------------------           GRAPH GAP Trumpler-14       ------------------------
%%-----------------------------------------------------------------------
%
%--------------
%%-----------------------------------------------------------------------
%%--------------------           GRAPH Sky Trumpler-14       ------------------------
%%-----------------------------------------------------------------------
%
%--------------

As discussed in section~\ref{sectionMethod}, our reason for using the sHRD instead of the classical HR is the lack of reliable distance measurements for all the stars analyzed in the sample to derive luminosities. The idea of the sHR diagram (presented in \citeauthor{Langer2014} \citeyear{Langer2014}) is to replace the luminosity (L) with the quantity $\mathcal{L}=T^{4}_{\rm eff}/g_{\rm true}$, which is the inverse of the flux-weighted gravity introduced by \cite{Kudritzki2003}.
It depends only on variables that can be directly derived from stellar spectra without knowledge of the stellar distance or the extinction, and it presents horizontal stellar evolutionary tracks for massive stars.
The differences and possible caveats between these two diagrams are discussed in \cite{Langer2014} or \cite{Castro2014}, but for this particular topic of stars near the ZAMS, the results are analogous, as can be seen in studies using the HR diagram with similar results \citep{Sabin-Sanjulian2017,Schneider2018}. It is possible, however, that if gravity measures are systematically inaccurate, the gap would only appear for the sHRD, but these cases should be studied individually, and a rather large systematic error ($\sim$ 0.25 dex) would be necessary to reconcile models and observations.

Here, we evaluate the possibility that the discrepancy only appears in the sHRD and disappears when we build the classic HR diagram. An extensive discussion of sHRD versus HRD can be found in \cite{Markova2018}. It is necessary then to obtain a distance measurement, and to derive the luminosity for each star. In this case, we decided to estimate a single distance to the group of stars that make up one cluster, Trumpler-14. 
The strength of this method is that the combination of the parallax values for all the available stars of a cluster eliminates the possible dispersion or even some systematic problems. For example, \textit{Gaia} parallaxes present problems for bright stars, which many O stars are \citep{Luri2018}. On the other hand, it is important to confirm that the analyzed stars really belong to this cluster, so that the combination of values is valid. For Trumpler-14, around 100 stars with \textit{Gaia} data were used to determine a parallax of 0.42 mas (see Appendix~\ref{AppTr14} and \citeauthor{Luri2018}\citeyear{Luri2018}), based on a method analogous to the work of \cite{Davies2019}, including the correction for the systematic offset of $-0.03$ mas \citep{GaiaCollaboration2018a}. 
Using this value, we calculated absolute magnitudes and then luminosities of the O stars of our sample belonging to the cluster, following standard procedures \citep{Kudritzki1980,Herrero1992,Repolust2004}. Then we combined the radius with the temperature to obtain the luminosity. Fig.~\ref{VGAP_Trumpler} shows the 8 stars in Trumpler-14 for which we have results in the HR diagram. The error bars shown are the formal error of our analysis, which greatly underestimate the actual error. A more realistic error, based on the distance uncertainty, would be around $\log{\textrm{L}}\pm0.05$ [dex].

Analyzing the result more in depth, we note that the position (on the Y-axis) of half the stars has varied. Very hot stars have moved very high or low from their original position in the sHRD, but they still play their role as very young stars in the sample. HD~93128 is now well within the gap area; however, we recall that the parameters obtained for stars with such a high \Teff\ are not properly constrained with our method (due to the lack of \ioni{He}{i} lines). This result indicates that the shape of the gap region must be taken with care when the region above the 60\msol\ evolutionary track is considered because their limits are still not precisely defined. Two stars of intermediate spectral type (O6) have moved below their original position, approaching the ZAMS and moving from the 2-3 Myr interval to the 1-2 Myr range. At the same time, their apparent youth has increased, they have moved to the part of the diagram where there was less discrepancy between models and observations. They have not entered the gap region. At first glance, it might be concluded that the problem has not been corrected when the HR diagram that includes the O-type stars in Trumpler-14 is inspected; however, a closer inspection of Fig.~\ref{VGAP_Trumpler} indicates that this is not the best sample to extract any firm conclusion in these regards. All considered stars have  effective temperatures higher than $\sim$45000~K or lower than $\sim$38000~K, which means that the intermediate range in \Teff\ where the gap region is mainly located is missing. Because switching from the sHRD to the HRD only affects the Y-axis, Trumpler-14 cannot be used to evaluate the possible effect on using the HRD instead of the sHRD when the presence of mid O-type stars close to the ZAMS is investigated.

\subsubsection{Role of a stochastic IMF in Trumpler-14 and the whole sample}

The particular case of studying an isolated cluster could be biased due to the stochastic nature of the initial mass function (IMF) in the formation 
of massive stars, or even due to the ill fortune of not having observed a pair of stars belonging to the cluster that could cover the void region. This bias disappears when the original complete sample of O stars is considered, but we decided to conduct the particular study in Trumpler-14 because it is one of the youngest clusters to which our survey has access.

In a first step, we evaluated the completeness of our sample of massive Trumpler-14 stars. We assumed a mass of Trumpler-14 close to $4 \cdot 10^3$ \msol\ \citep{Sana2010, Hur2012,Alexander2016} and a Kroupa IMF  \citep{Kroupa2001}. The mean mass per star of this IMF is $0.357$ \msol\ (with a mass limit between 120--0.01 \msol), which translates into $N_* \sim 11 203$ stars in the cluster for the given cluster mass. This IMF also provides a probability that the mass of a random star is in the $25 - 60$ \msol\ mass range of $p_{[25,60]} = 5.14 \cdot 10^{-4}$, and $p_{[60,120]} = 1.38 \cdot 10^{-4}$ for the $60 - 120$ \msol\ mass range. When we use the 11 203 stars and the previous probabilities in a multinomial distribution, the expected number of stars in the $25 - 120$ \msol\ mass range is $7\pm 4$ at a 90\% confidence interval, which is consistent with our census of 10 stars in this mass range. More in detail, the probability of finding exactly 7 stars is 14.8\%, the probability of finding exactly 10 stars is 8\%, and finally, the probability that the cluster would have more than 12 stars in this mass range is 3.6\%. Accordingly, the probability of finding exactly 13 stars is 1.8\%. In summary, for the assumed mass and IMF, we are therefore relatively confident that our star census is complete.

In a second step, we investigated how likely a configuration would be for which no star is found in the $25 - 60$ \msol\ mass range and three stars appear in the $60 - 120$ mass range. This would correspond to a two-bursts scenario where all the lower mass stars were formed more than 2 Myr ago, and the three more massive stars observed were formed in the last 2 Myr. We assumed that each of these two bursts formed half of the stars each ($\pm1$ because $N_*$ is a even number). 
In this case, the probability\footnote{This probability can be obtained directly by the Bayes theorem over the associated multinomial distribution. For the particular case of the probability of $n_{[25-60]} = 0$ given that $n_{[60-120]} = n_a$ is
\begin{equation}
P(n_{[25-60]} = 0\, |\, n_{[60-120]} = n_a; N_*) = \left( \frac{1-p_{[25-60]}-p_{[60-120]}}{1-p_{[25-60]}}\right)^{N_* - n_a}.
\end{equation} Similarly for other cases as $P(n_{[60-120]} = 0\, |\, n_{[25-60]} = n_a;N_*)$  when the corresponding changes have been made.} of having zero stars in the $60 - 120$ \msol\ mass range given that there are seven stars in the $25 - 60$ \msol\ mass range is 46\%. On the other hand, the probability of having zero stars in the $25 - 60$ \msol\ mass range given that there are three stars in the $60 - 120$ \msol\ mass range is 5.7\%. With this value, although it is low, we cannot discard the possibility that the gap observed in Trumpler-14 is due to IMF sampling effects. It is highly improbable, but plausible. However, we note that these oddities increase when the number of stars considered increases. As an example, the probability of having zero stars in the $25 - 60$ \msol\ mass range given 3 stars in the $60 - 120$ \msol\ mass range and $N_* = 11 203$ drops to 0.3\%. This situation would correspond to the case where Trumpler-14 and Trumpler-16 are considered as a whole (this would require a study of Trumpler-16, which is beyond of the scope of this paper). Finally, the situation would become even more improbable when the overall IACOB sample is taken into account, but again, a study like this would require evaluating the different star formation history of all the clusters included in the sample.

%%%%%%%%%%%%%%%%%%%%%%%
%%%%%%%%%%%%%%%%%%%%%%%%%%%%%%%%%%%%%%%%%%%%%%%%%%%%%%%%%%%%%%%%%%%%%%%%%
\subsection{Some insights from stellar evolution modeling}\label{sectionOri}
%%%%%%%%%%%%%%%%%%%%%%%%%%%%%%%%%%%%%%%%%%%%%%%%%%%%%%%%%%%%%%%%%%%%%%%%%

%%%%%%%%%%%%%%%%%%%%%%%%%%%%%%%%%%%%%%%%%%%%%%%%%%%%%%%%%%%%%%%%%%%%%%%%%
\subsubsection{ZAMS and theoretical birthline of massive stars}\label{SectAccret}
%%%%%%%%%%%%%%%%%%%%%%%%%%%%%%%%%%%%%%%%%%%%%%%%%%%%%%%%%%%%%%%%%%%%%%%%%

\begin{figure}
\includegraphics[width=0.49\textwidth]{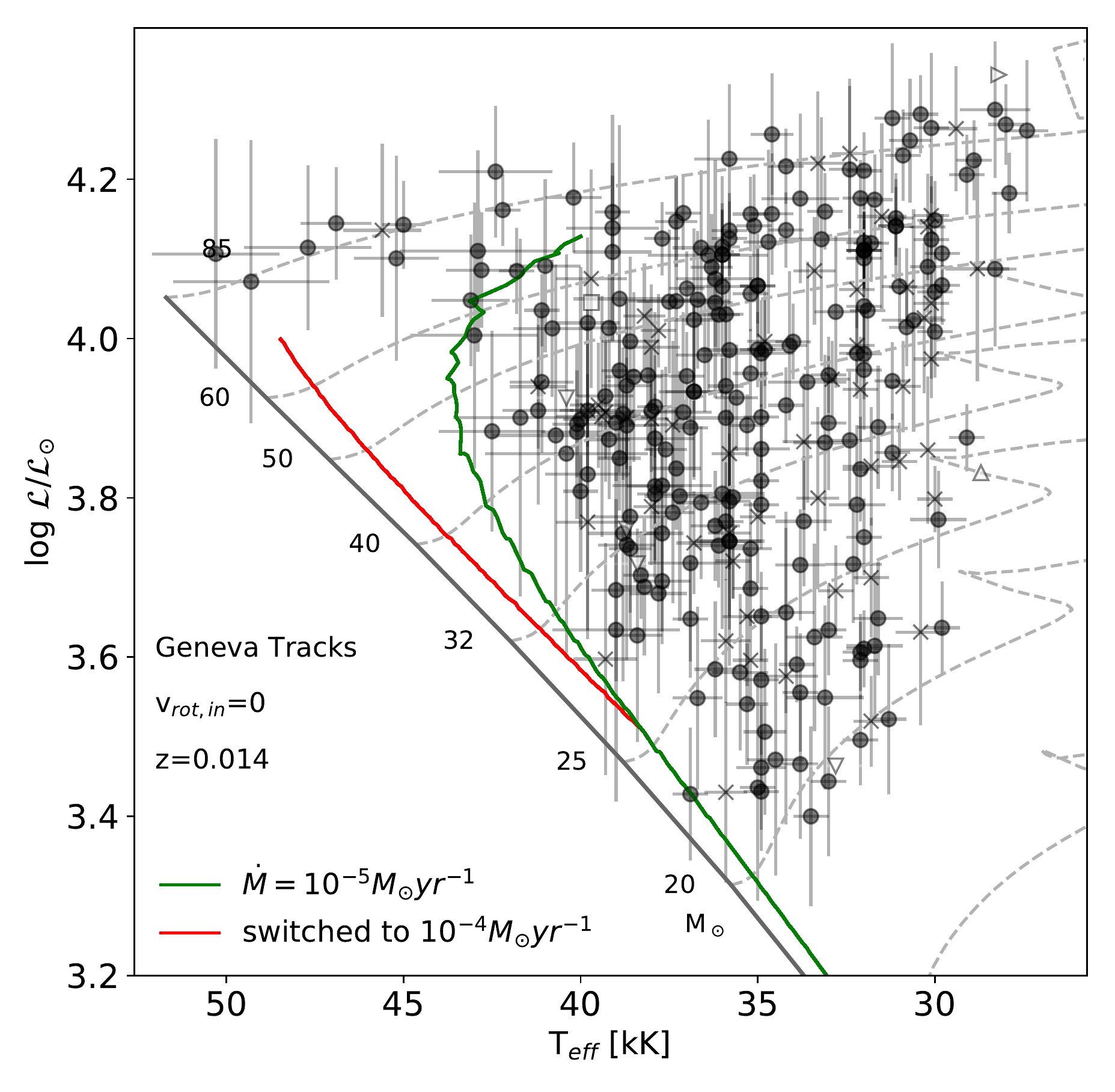}
\caption{Same as Fig.~\ref{VGAP}. Additional evolutionary tracks of the model evolving at mass accretion $\dot M=10^{-5}\rm\,M_\odot\,yr^{-1}$ (solid green). The red track corresponds to the model in which the evolution is switched from $\dot M=10^{-5}\rm\,M_\odot\,yr^{-1}$ to $10^{-4}\rm\,M_\odot\,yr^{-1}$ at $M=25\rm\,M_\odot$.}
\label{hr5}
\end{figure}

\begin{figure}
\includegraphics[width=0.49\textwidth]{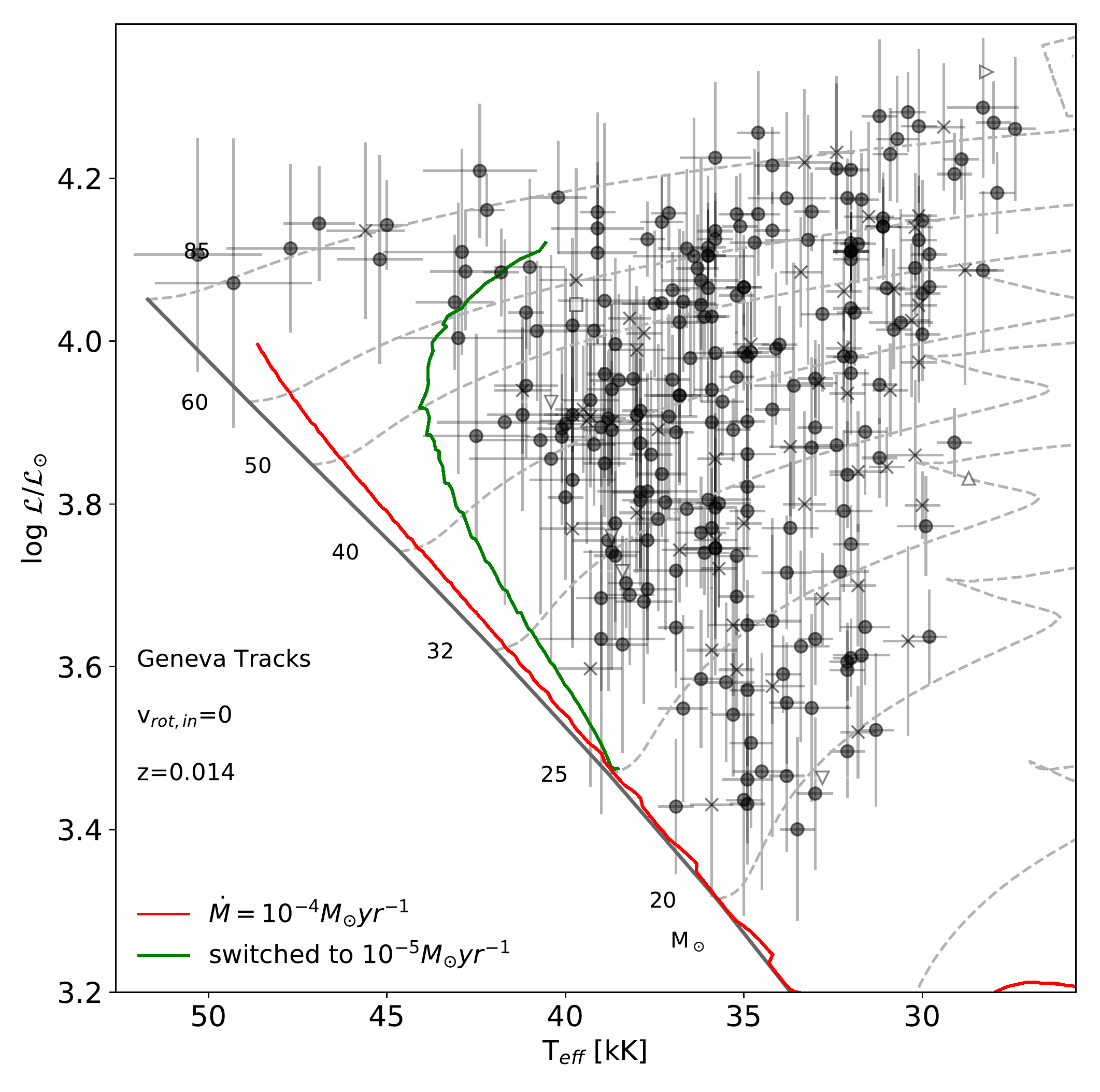}
\caption{Same as Fig.~\ref{hr5} for $\dot M=10^{-4}\rm\,M_\odot\,yr^{-1}$ (solid red).
The green track corresponds to the model in which the evolution is switched from $\dot M=10^{-4}\rm\,M_\odot\,yr^{-1}$ to $10^{-5}\rm\,M_\odot\,yr^{-1}$ at $M=25\rm\,M_\odot$.}
\label{hr4}
\end{figure}

\begin{figure}
\includegraphics[width=0.49\textwidth]{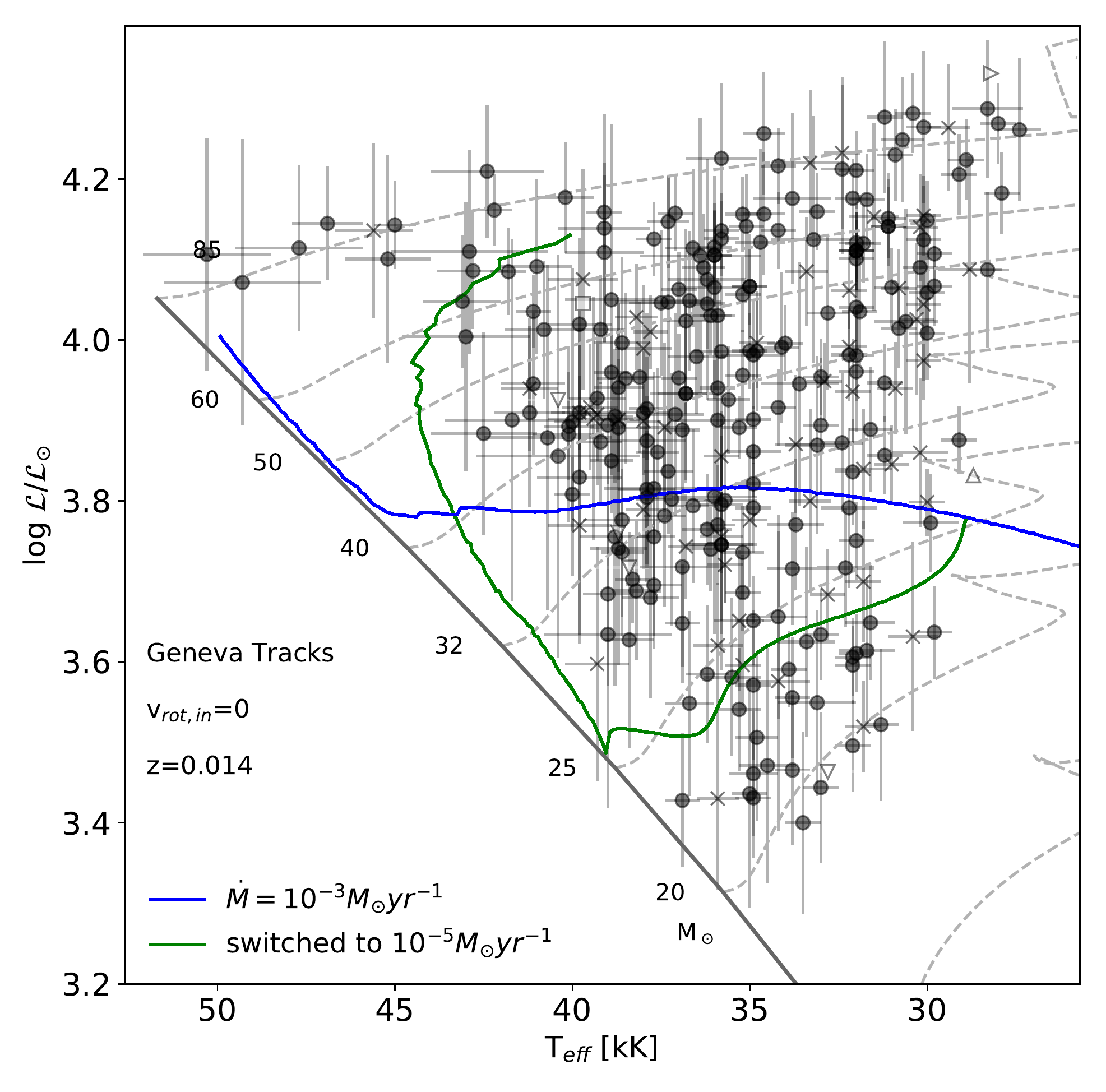}
\caption{Same as Figs.~\ref{hr5} and \ref{hr4} for $\dot M=10^{-3}\rm\,M_\odot\,yr^{-1}$ (solid blue).
The green track, which begins at the blue line at $M=25\rm\,M_\odot$, corresponds to the model in which the evolution is switched from $\dot M=10^{-4}\rm\,M_\odot\,yr^{-1}$ to $10^{-5}\rm\,M_\odot\,yr^{-1}$ at $M=25\rm\,M_\odot$.}
\label{hr3}
\end{figure}

We described in Sect.~\ref{section1} that the formation path of a star is defined as the birthline, which in massive stars (with standard accretion rates) could pass through stellar fusion ignition (ZAMS) before reaching complete accretion of its birth cloud. For a fixed accretion rate, all massive stars follow the same birthline until accretion stops, reaching their maximum masses and continuing then through a canonical evolutionary track.
The position in the HR (or sHR) diagram of a star largely depends on the amount of hydrogen that has already been consumed in its core through fusion processes. For a slower accretion rate, the evolution of the star is also slower and allows enough time for hydrogen to be consumed. This would cause the star to move in the HR (or sHR) diagram in an upward right direction \citep[see e.g.,][]{Vanbeveren1998}. We determined whether it is possible to reconcile the two models (the birthline) and observations by adjusting or tuning their critical factor, the accretion history included in the models. The process for generating the models with different accretion parameters and their evolutionary tracks is described in Appendix~\ref{AppHaemm}. Here we only discuss the shape of the tracks in the region of the HR diagram that corresponds to the observations.
The tracks are shown on Figs.~\ref{hr5}, \ref{hr4}, and \ref{hr3}.

When the gap near the ZAMS in the OB range is to be interpreted as an effect of accretion,
the accretion track has to correspond to the left-hand side envelope of the observations, that is, the birthline\footnote{Observationally, the birthline is sometimes defined as the line on the HR diagram (or sHRD) along which young stars become visible in the optical (Fig. 1 in \citeauthor{haemmerle2019b}\citeyear{haemmerle2019b}). The birthline of massive stars is expected to correspond to the lower envelope of the observations \citep{haemmerle2019b}.}.
When a star on the accretion track reaches its final mass, its subsequent evolutionary track is toward the red, where observations show the presence of stars. This is in our case, when the accretion time is longer than the KH time.

Fig.~\ref{hr5} shows that the model simulations accreting at $\dot M=10^{-5}\rm\,M_\odot\,yr^{-1}$ match the envelope of the observations well, except for the series of very high-mass stars. As explained throughout this article, stars in this region are susceptible to large uncertainty due to the absence of \ioni{He}{i} diagnostic lines.
This model simulation reaches the ZAMS at a relatively low mass ($M\simeq8\rm\,M_\odot$, not visible in the plot).
After this, as accretion proceeds, the model approximately follows the ZAMS toward the blue until $M\simeq30\rm\,M_\odot$.
Then it shifts toward higher luminosities, and at $M\simeq50\rm\,M_\odot$, it starts to move significantly toward the red as a result of the MS evolution.
This can be understood by comparing the timescales for accretion ($t_{\rm accr}=M/\dot M$) and MS evolution ($t_{\rm MS}$).
For instance, at $M=10\rm\,M_\odot$, the former is $t_{\rm accr}=1$~Myr, more than one order of magnitude shorter than the latter
($t_{\rm MS}\simeq20$ Myr, \citealt{schaller1992}).
In this mass range, the star can therefore efficiently grow in mass before it evolves significantly on the MS.
As a consequence, its track remains close to the ZAMS as accretion proceeds.
For $M=30\rm\,M_\odot$, the two timescales become comparable ($t_{\rm accr}=3$ Myr and $t_{\rm MS}\simeq6$ Myr).
The effects of accretion and MS evolution both affect the evolutionary birthline track, which is progressively shifted from the ZAMS.
Finally, at $M=50\rm\,M_\odot$, the accretion time becomes longer than the MS time ($t_{\rm accr}=5$ Myr and $t_{\rm MS}\simeq4$ Myr),
the effect of MS evolution on the track dominates that of accretion, and the star moves away from the ZAMS.

The models at $10^{-4}\rm\,M_\odot\,yr^{-1}$ (Fig.~\ref{hr4}) and $10^{-3}\rm\,M_\odot\,yr^{-1}$ (Fig.~\ref{hr3}) show a different behavior.
These two models reach the ZAMS at higher masses, $M\simeq15\rm\,M_\odot$ and $40\rm\,M_\odot$ , respectively.
At these rates, the accretion time always remains shorter than 1 Myr until the final mass of the runs
($t_{\rm accr}<70\rm\,M_\odot/10^{-4}\,M_\odot\,yr^{-1}=0.7$ Myr), while the MS time always remains longer than 3 Myr.
Thus the evolutionary tracks are determined essentially by accretion, and almost follow the ZAMS as accretion proceeds.

The relevant accretion rate for estimating the accretion time is the current rate.
As a consequence, the models with an accretion rate that changes at $M=25\rm\,M_\odot$ do not keep memory of their past accretion history.
The tracks of these models are shown in Figs.~\ref{hr5}, \ref{hr4}, and \ref{hr3}.
A star with $M>25\rm\,M_\odot$ that accretes at $\dot M\geq10^{-4}\rm\,M_\odot\,yr^{-1}$ follows the ZAMS as accretion proceeds,
independently of the previous accretion history (red track in Fig.~\ref{hr5}).
In contrast, a star with $M>25\rm\,M_\odot$ accreting at $\dot M=10^{-5}\rm\,M_\odot\,yr^{-1}$ always evolves away from the ZAMS,
even if it accreted at $\dot M\geq10^{-4}\rm\,M_\odot\,yr^{-1}$ for $M<25\rm\,M_\odot$ (green tracks in Figs.~\ref{hr4} and \ref{hr3}).

\subsubsection{Constraints on accretion history from observations}\label{ConstrAcc}

In accordance with previous suspicions from \cite{Vanbeveren1998}, the models described in the previous section show that for the birthline to fit the envelope of the observations,
the accretion rate must not exceed $10^{-5}\rm\,M_\odot\,yr^{-1}$ in the mass range $M\gtrsim25\rm\,M_\odot$.
This result is counterintuitive because for massive stars, typical accretion rates of $10^{-3}\rm\,M_\odot\,yr^{-1}$ are expected \citep{larson1971,hosokawa2009,kuiper2010b}.

This suggests that a typical accretion history given by
\begin{equation}
\dot M=\left\{\begin{array}{l}
10^{-3}\,-\,10^{-5}{\rm\,M_\odot\,yr^{-1}\quad for\  }M\lesssim25\rm\,M_\odot\\
10^{-5}{\rm\,M_\odot\,yr^{-1}\quad for\ }M\gtrsim25\rm\,M_\odot
\end{array}\right..
\end{equation}

The accretion history of massive stars remains an open question.
Hydrodynamical simulations of the pre-stellar collapse show a rich variety of behaviors, with rates that increase or decrease with time, depending essentially on the initial conditions \citep[e.g.,][]{peters2010,peters2010a,peters2010b,peters2011,kuiper2010,kuiper2011,girichidis2011,girichidis2012,girichidis2012a,meyer2018}.
Observations of outflows around massive young stellar objects (or MYSOs) and of the distribution of Herbig Ae/Be stars on the HR diagram suggest a rate that increases with the stellar mass \citep{Behrend2001,haemmerle2019b}.
This behavior is supported by the luminosity distribution of massive protostars in the Milky Way \citep{davies2011}.
On the other hand, the probability dependence of the rate on the stellar mass does not necessarily reflect an evolutionary sequence. More massive objects might form at higher rates during the main accretion phase, before the rate declines.
Many reasons might explain the decrease in the accretion rate as a function of the mass of the accretor,
such as the effect of the rising UV feedback \citep{peters2010b} as the stellar surface heats up, the increase of radiation pressure \citep{kuiper2010},
or the angular momentum barrier \citep{haemmerle2017}.

%%%%%%%%%%%%%%%%%%%%%%%%%

\subsubsection{Very massive stars in the sample}\label{VMS}

Our sample of 285 O-type stars with spectroscopic parameters includes a small group of 6 early O-type stars with evolutionary masses $\sim$70~\msol\ that stands out (see, e.g.,  Fig.~\ref{VGAP}). These stars do not fit in the birthline constructed by considering a mass accretion rate in the star formation phase one order of magnitude lower than traditionally considered (see the green and red solid lines in Figs.~\ref{hr5} and \ref{hr4}, which correspond to 10$^{-5}$ and 10$^{-4}$~$M_\odot\,yr^{-1}$, respectively). Beyond considering this result as a counterargument against this alternative scenario of massive star formation, it might also be interpreted as empirical evidence that these stars have followed a different (nonstandard) evolution. Some possibilities that have been proposed are binary interaction in general, mergers in particular, and/or chemically homogeneous evolution. 

The first possibility is supported by recent findings indicating that a large portion of massive stars are born as part of a binary (and multiple) system. In most of the cases, binary interaction is expected to critically affect their evolution \citep{Sana2012,Wang2020}.

Binary-evolution models predict that a large portion of close binaries may exchange mass before they leave the MS \cite[e.g.,][]{deMink2013}, with cases where the secondary mass has been increased, which produces a brighter star than any of the pre-interaction stars at a coeval age. This mass-transfer effect might led to a secondary sample of stars that appear rejuvenated and with higher surface temperatures than the remaining stars that belong to the same cluster \citep{Schneider2014,Wang2020}.

Another possibility is that in extreme cases, some interacting binary systems might evolve into a configuration where both stars overfill their Roche lobes and eventually produce a merger \citep{Benson1970,Wellstein2001,Pols1994,Wellstein2001, deMink2007,deMink2012,deMink2014}.
On average, the percentage of mergers expected to occur in a given population of massive MS stars has been estimated to stay in the order of 10\% \citep{Podsiadlowski1992,Eldridge2011}.

Stars resulting from a merger are not only more massive and more luminous than any of the stars in the progenitor binary system \citep{deMink2013,Schneider2014,Schneider2019}, but they are also predicted to suffer from a rejuvenation process as a result of renewed content of fresh hydrogen gas that is mixed into the central burning regions \citep{Glebbeek2013,deMink2014}. In this sense, mergers are expected to be observationally detected as blue stragglers, that is, stars that appear to be younger than the age of the cluster in which they reside and are therefore closer to the ZAMS
\citep[e.g.,][]{Sills2002,Glebbeek2008,Mermilliod1982,Chen2009,Lu2010}.

The two channels indicated above might help to explain the presence of stars in the top left region of our empirical sHRD in the context of the lower accretion rate scenario; however, it might be argued than the same process is expected to lead to a similar filling of the gap region. One possible explanation to overcome this caveat might be that as indicated by \cite{Langer2019}, only the stars above a certain mass have a phase where the envelope inflates and leads to a merger instead of the star-star mass transfer. In this context, it is also interesting to note that the relative percentage of SB2 systems identified in our sample of 160 O dwarfs (see Fig.~\ref{Hist_Comp_GOSC_EBV_V}) in which the primary component has a mid-O spectral type is much higher than when the primary is an early O-type star. While not necessarily a definitive solution, we consider that this line of argument deserves a more in-depth study in the future.

An alternative or complementary explanation is the so-called chemically homogeneous evolution. Massive stars that reach a very high rotation rate (either during the star formation process or by angular momentum transfer in a binary system) are predicted to follow a peculiar evolution, evolving left- and upward in the HR diagram \citep{Maeder1987,DeMink2010,Brott2011}. 

\cite{Martins2013} showed empirical evidence of chemically homogeneous evolution taking place in environments with metallicity up to solar; however, this study was based on the investigation of a sample of early-type H-rich WN stars. Although several authors have reported observational support for it in subsolar metallicity \citep{Bouret2003,Walborn2004,Bouret2013}, these studies are just based on a few targets, and similar findings have not been reported yet at solar metallicity.
While more work is also needed in this direction (including the identification of peculiar nitrogen and helium surface abundance patterns in the very early O-type stars, see e.g., \citeauthor{RiveroGonzalez2012} \citeyear{RiveroGonzalez2012}), it is interesting to note that the early-O stars in our sample do not have extreme values of \vsini\ \citep{Holgado2019}, as is required by massive stars to follow this type of evolution \citep{Maeder1987}.

%%%%%%%%%%%%%%%%%%%%%%%%%%%%%%%%%%%%%%%%%%%%%%%%%%%%%%%%%%%%%%%%%%%%%%%%%%%%%%%%%
\section{Summary and conclusions}\label{sectionConcl}

The main conclusion that can be highlighted from this study is that {\em \textup{the lack of O-type stars detectable in optical wavelengths and with parameters compatible with the theoretical ZAMS is a real and robust empirical fact.}}
We summarize below the main line of work we have followed to reach this conclusion, and some other implications of this finding.

We performed a spectroscopic study of a sample of 415 Galactic O-type stars, implying 5-10 times stars more than any previous similar study in the literature. We showed that this sample is a good representation, in terms of spectral type and luminosity class, of the list of stars included in version 4.1 of the Galactic O star catalog, comprising $\sim$70\% of the stars quoted there. 
For most luminosity classes, the completeness (with respect to GOSC\,v4.1) approaches 75\%. For the dwarf sample, this number is slightly smaller (60\%). We demonstrated that our sample is not significantly affected by any systematic observational bias with respect to the GOSC sample (in particular, regarding stars with mid-O spectral types, corresponding to the region where the lack of stars close to the ZAMS is more pronounced). 
We consider that the sample is not clearly biased either for stars that are relatively extinguished by surrounding material from the associated parental cloud, although additional exploration is possible.
We therefore suggest that any further attempt to confirm the existence of stars that are located closer to the theoretical ZAMS (in the mid-O spectral type range) than our sample of stars must imply the use of infrared observations, which can penetrate thicker layers of material and dust.

We performed a homogeneous quantitative spectroscopic analysis of 285 of the 415 stars in our initial sample (including likely single and SB1 stars, and excluding SB2 systems and a few stars with peculiar spectroscopic features), and located them in the Kiel and spectroscopic HR diagrams. We also collected and presented in the sHR diagram information from another ten studies from the literature performing quantitative spectroscopy of Galactic O-type stars, based on different methods and stellar atmosphere codes. The shortage of O stars near the ZAMS prevails, and therefore we discarded any particular deficiency of our analysis strategy as the origin of the mismatch between observations and theory. A potential, more general, shortcoming in current stellar atmosphere codes, related to not considering the effect of turbulent pressure in the computation of the stellar atmosphere structure, and which could help to populate the gap region, was also briefly discussed.

We performed a more in-depth study of one of the youngest known Galactic clusters for which we have available high-resolution spectra: Trumpler-14. Compared with the GOSC catalog, we find that we do not lack decisive stars to cover the unpopulated area. We then constructed the HR diagram of the Trumpler-14 sample using \textit{Gaia}-DR2 data. We evaluated the distribution of the sample and compared it with the analogous sHR diagram that we used. The general results are similar in both diagrams: equivalent area covered, and a lack of mid-O-type stars near the ZAMS. 
In addition, when we assumed a mass for the cluster as inferred from previous studies,  the number of O stars that we assigned to Trumpler-14 is practically complete, and a large number of additional O stars belonging to that cluster are not expected.

The empirical evidence presented here for the existence of a gap close to the theoretical ZAMS
represents (if not refuted by future observations) an important challenge for our current theories about the formation and evolution of massive stars. When we expand on this idea, in the case of high-mass stars, the theoretical ZAMS (which is based on instantaneous hydrogen ignition at a given mass) might be a different concept from the observed ZAMS, which is located at cooler effective temperatures and is related to the accretion timescale (i.e., the mass-accretion rate). In this sense, the hotter envelope of the observed distribution of O-type stars might trace a type of stellar birthline, as in the case of cooler stars, but this time, after starting the ignition of hydrogen in their cores.

In this line of argument, the variation in mass-accretion rate during star formation is a parameter that might help to explain the lack of empirically detected O-type stars close to the theoretical ZAMS. While the commonly assumed hypothesis of an accretion rate that is constant or increases with the mass of the accretor is not able to reproduce the empirical data, we showed that a decreasing rate of accretion with mass (turning from $\sim10^{-4}$ to $\sim10^{-5}$~\msol/year at $\sim$25~\msol) might reconcile observations, evolutionary models, and hydrodynamical simulations.

A possible caveat to this proposal are the few O2\,--\,O3.5 dwarfs that are found much closer to the ZAMS than the remaining stars in our sample. However, this result might also be interpreted as empirical evidence indicating that this population of early O-type stars have experienced binary interaction, are the products of a stellar merger event, and/or are following a chemically homogeneous evolution. Although we cannot yet reach firm conclusions with the information presented in this paper, we propose that this line of argument deserves a more in-depth study in the future. In these regards, including in the investigated sample multi-epoch observations of all known early-type stars\footnote{For example, the $\sim$15 stars in NGC~3603 with spectral types in the range O3\,--\,O3.5 identified by \cite{Melena2008}, which are not included in our current sample due to their magnitude.} will allow us to obtain more statistically significant conclusions.

%%%%%%%%%%%%%%%%%%%%%%%%%%%%%%%%%%%%%%%%%%%%%%%%%%%%%%%%%%%%%%%%%%%%%%%%%%%%%%%%%
\begin{acknowledgements}
%%%%%%%%%%%%%%%%%%%%%%%%%%%%%%%%%%%%%%%%%%%%%%%%%%%%%%%%%%%%%%%%%%%%%%%%%%%%%%%%%

We express our extreme gratitude to the referee for the time devoted as their comments really contributed to improve this work. GH wants to thank J.Puls for his helpful comments and precise corrections that improved this manuscript significantly. GH acknowledges that this research has been partially funded by the Spanish State Research Agency (AEI) Project No. ESP2017-87676-C5-1-R and No. MDM-2017-0737 Unidad de Excelencia “María de Maeztu”- Centro de Astrobiología (INTA-CSIC). S-SD, DJL, and AHD acknowledge support from the Spanish Government Ministerio de Ciencia, Innovaci\'on y Universidades through grants PGC-2018-091\,3741-B-C22, and from the Canarian Agency for Research, Innovation and Information Society (ACIISI), of the Canary Islands Government, and the European Regional Development Fund (ERDF), under grant with reference ProID2017010115. LH and GM thanks the Swiss National Science Foundation (project number 200020-172505). GM has received funding from the European Research Council (ERC) under the European Union's Horizon 2020 research and innovation program (grant agreement No 833925).
Based on observations made with the Nordic Optical Telescope, operated by NOTSA, and the Mercator Telescope, operated by the Flemish Community, both at the Observatorio del Roque de los Muchachos (La Palma, Spain) of the Instituto de Astrof\'isica de Canarias. Based on observations at the European Southern Observatory in programs 073.D-0609(A), 077.B-0348(A), 079.D-0564(A), 079.D-0564(C), 081.D-2008(A), 081.D-2008(B), 083.D-0589(A), 083.D-0589(B), 086.D-0997(A), 086.D-0997(B), 087.D-0946(A), 089.D-0975(A).
This work has made use of data from the European Space Agency (ESA) mission {\it Gaia} ({\tt https://www.cosmos.esa.int/gaia}), processed by the {\it Gaia} Data Processing and Analysis Consortium (DPAC, {\tt https://www.cosmos.esa.int/web/gaia/dpac/consortium}). Funding for the DPAC has been provided by national institutions, in particular the institutions participating in the {\it Gaia} Multilateral Agreement.

%%%%%%%%%%%%%%%%%%%%%%%%%%%%%%%%%%%%%%%%%%%%%%%%%%%%%%%%%%%%%%%%%%%%%%%%%%%%%%%%%
\end{acknowledgements}
%%%%%%%%%%%%%%%%%%%%%%%%%%%%%%%%%%%%%%%%%%%%%%%%%%%%%%%%%%%%%%%%%%%%%%%%%%%%%%%%%

%%%%%%%%%%%%%%%%%%%%%%%%%%%%%%%%%%%%%%%%%%%%%%%%%%%%%%%%%%%%%%%%%%%%%%%%%%%%%%%%%
\bibliographystyle{aa}
\bibliography{sample}

%%%%%%%%%%%%%%%%%%%%%%%%%%%%%%%%%%%%%%%%%%%%%%%%%%%%%%%%%%%%%%%%%%%%%%%%%%%%%%%%%

%\newpage

\appendix

\section{HD199579: an O6.5\,Vz star within the gap region?}\label{AppHD199579}

\begin{figure}[t]
\centering
\includegraphics[width=0.45\textwidth]{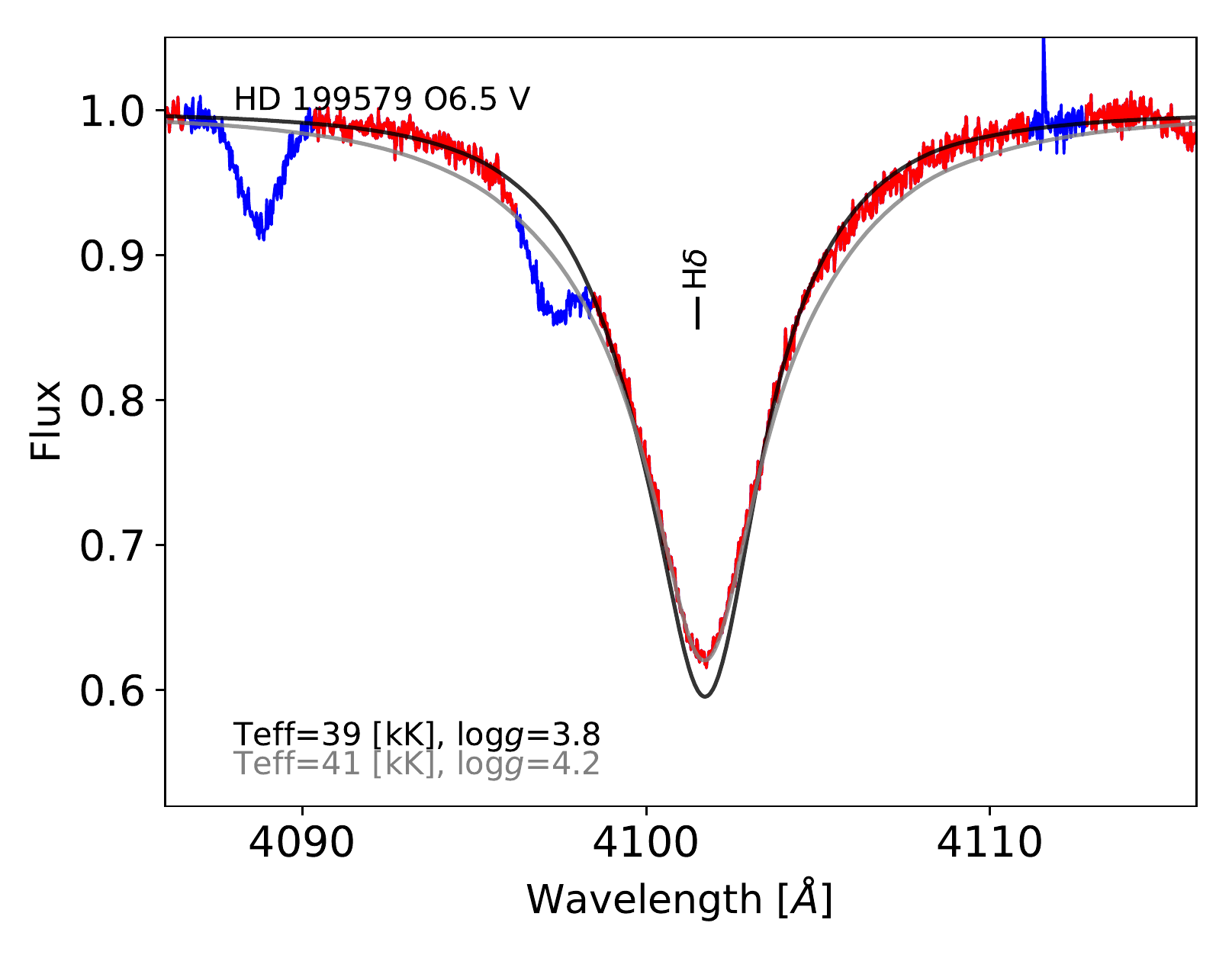}
\includegraphics[width=0.45\textwidth]{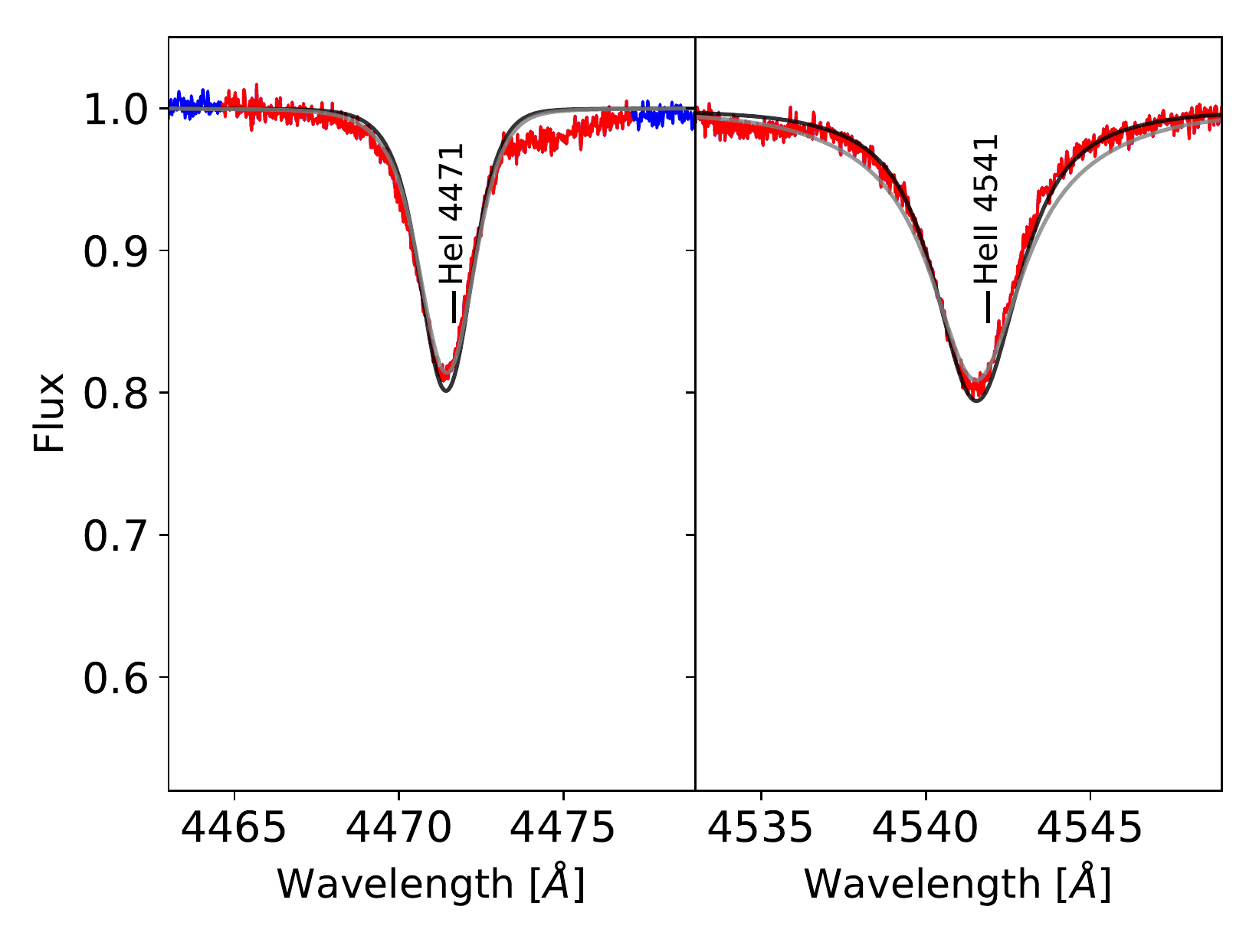}
\includegraphics[width=0.45\textwidth]{./Crop_xi2_VH.png}
\caption{\textit{Top and middle} Comparison of the synthetic spectra of two best-fitting {\sc fastwind} models to the observed spectrum of the O6.5~Vz star HD~199579 for three diagnostic lines. The first best-fitting model corresponds to the {\sc iacob-gbat} analysis with \Teff\ and \grav\ as free parameters (the red part is fitted and the blue part is ignored). In the second model, \grav\ was fixed to 4.2~dex. \textit{Bottom} $\chi^2$ distributions for \Teff\ and \grav\ resulting from the {\sc iacob-gbat} analysis. The horizontal dashed lines represent the value of $\chi^2$ for the best-fitting model (red dots) and the 1$\sigma$ and 2$\sigma$ confidence levels. Any model with \grav$\sim$4.2 is clearly beyond the 2$\sigma$ level.}
\label{Forced42_VH}
\end{figure}

The only star that is located within the gap region in Fig.~\ref{VGAP_Lit} \citep[according to the parameters determined by][]{Martins2015} is HD~199579, a double-line spectroscopic binary including a dominant O6.5~Vz main component and a very faint companion \citep{Williams2001}. \citeauthor{Martins2015} performed a spectroscopic analysis of a high-resolution optical spectrum of the star with the {\sc cmfgen} code \citep{Hillier1998,Hillier2001,Hillier2012} and found \Teff\,=\,41\,500 K, \gravt\,=\,4.15 dex). Our analysis for this star provides quite different values (\Teff=39.5$\pm$0.8 kK, \gravt=3.9$\pm$0.1 dex), however. As an exercise to evaluate to which extent the parameters provided by \citeauthor{Martins2015} can be considered a valid solution from the {\sc iacob-gbat/fastwind} analysis, we launched the tool again, but forcing the surface gravity to a value closer to that expected in the ZAMS \citep[and close to the one obtained by][]{Martins2015}, that is, \grav\,=\,4.2~dex. The results are shown in Fig~\ref{Forced42_VH} and prove that when the surface gravity is forced to a higher value, the associate \fastwind\ best-fitting model is not able to fit the wings of the H$\delta$ and \ioni{He}{ii}\,$\lambda$4541 lines accurately. Further inspection of the bottom panel of Fig.~\ref{Forced42_VH} indicates that this set of parameters obtained by \citeauthor{Martins2015} is well above the 3$\sigma$ tolerance level of the $\chi^2$-fitting. We note that the difference in surface gravity is beyond the $\sim$0.10\,--\,0.15 dex systematic difference that has been claimed to be present between the {\sc fastwind} and {\sc cmfgen} codes \citep{Massey2013, Holgado2018}.

\begin{figure}[t]
\centering
\includegraphics[width=0.5\textwidth]{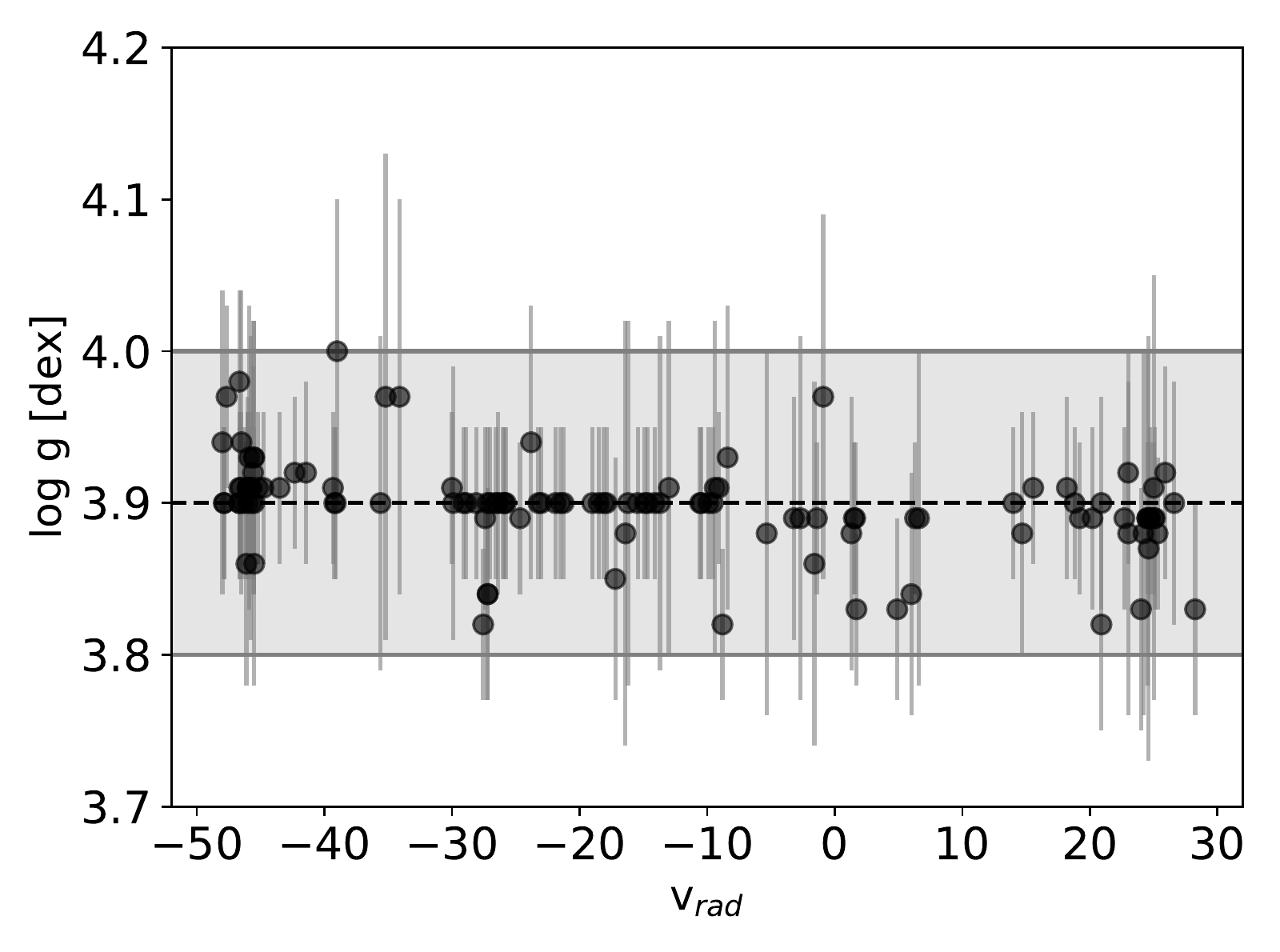}
\caption{Results from the {\sc iacob-gbat} analysis (regarding \grav) of 119 multi-epoch spectra of the SB2 star HD~199579, sorted according to the radial velocity of the spectrum. Uncertainties from the analyses are included as vertical error bars. The uncertainty in the original analysis in \grav\ ($\pm$0.1 dex) is shown as a gray area around the value obtained there, which is also the median value of all estimates (\grav\,=\,3.9~dex).}
\label{Grav_Black}
\end{figure}

Another possible explanation for the disagreement in the derived parameters of HD~199579 is that we did not analyze the same spectrum as \cite{Martins2015}. The multi-epoch character of the IACOB survey allows us to perform an academic exercise to check this possibility by analyzing the 119 available spectra. The resulting values and uncertainties are presented in Fig.~\ref{Grav_Black}, sorted according to the radial velocity measurement of each spectrum (we recall that HD~199579 is an SB2 system in which the secondary is much fainter than the primary, and that it is therefore not easily detected in the individual spectra). All our measurements are within the grid-step in \grav\ of $\pm$0.1 dex, and far away from the gravity estimated by \cite{Martins2015} (and the one expected for the star to be on the ZAMS).

Therefore it is not evident that the result of \cite{Martins2015} for HD~199579 can be considered as a critical counterexample of the absence of O-type stars near the ZAMS. Even considering the parameters derived by \citet{Martins2015}, given the fact that this is the only star located within the gap, the scarcity of stars in this region is still a remarkable empirical result. 

%%-----------------------------------------------------------------------
%%--------------------           GRAPH ForzGrav       ------------------------
%%-----------------------------------------------------------------------
%
%--------------

%%-----------------------------------------------------------------------
%%--------------------           GRAPH grav range       ------------------------
%%-----------------------------------------------------------------------
%
%--------------

%%-----------------------------------------------------------------------

\section{Membership and distance of Trumpler-14}\label{AppTr14}

Based on the discussion in Sect. \ref{EmpInsi}, it is clear that due to its extreme youth, $\leq$1 Myr, Trumpler-14 is critical in verifying the reality of the observed gap in the sHRD because the presence or absence of mid-O dwarfs close to the ZAMS in such a young cluster is a very strong constraint.  
Determining cluster membership of Trumpler-14 is therefore extremely important, and while the distribution of stars in the sHRD does not depend on distance, this is relevant for deciding whether individual stars are cluster members.
As pointed out in Sect. \ref{EmpInsi}, the mutual proximity of the clusters Trumpler-14 and Trumpler-16 has resulted in some confusion regarding cluster membership. For example, the stars or stellar systems HD\,93161, HD\,93160 and HD\,93250 are sometimes attributed to Trumpler-16 even though they are close to Trumpler-14, see, for example, the discussion in \citet{walborncarina} and \citet{smith2006}.
However, as noted in these papers (see also \citet{Hur2012}), these clusters have very similar distances such that separating individual stars on the basis of their distance would appear to be difficult.
We therefore used \textit{Gaia}-DR2 data \citep{Lindegren2018} to investigate cluster membership probabilities based on distance and proper motion, as discussed below.

We extracted all $\sim$150\,000 sources within 30$\arcmin$ of the center of Trumpler-16. This region includes Trumpler-14 and also Trumpler-15, a rather older cluster several arcminutes north of Trumpler-14.
The region also hosts stars belonging to other potential comoving groups such as  Collinder 228 and 232. Because of their sparsity and the debate about their reality as physical groups, discussed by \cite{Walborn1973} and \cite{smith2006}, we do not discuss these further in this paper.
We applied the following filters to these sources: \texttt{pmra\_error}$<0.07$ mas/yr, \texttt{pmdec\_error}$<0.07$ mas/yr, $0.3<$\texttt{parallax}$<0.5$ mas, and \texttt{bp\_rp}$<1.2$. The restrictions on proper motion are prompted by the discussion of outliers in the crowded regions of 30\,Dor by \citet{lennon2018} \citep[see also][]{platais2018}, while the color cut was used to exclude red giants. The color restriction also excludes massive stars that have intrinsic high extinction (known to exist in the region), but we note that the objective here is not to derive a complete list of cluster members, but rather to define secure cluster members. In this context, for the clusters in question, we defined as candidates all sources within 5$\arcmin$ of the cluster centers (as defined in Table \ref{pmtable}) and removed outliers in both proper motion directions with a 2$\sigma$ clipping algorithm. The results for the mean parallax and proper motion quantities are listed in Table \ref{pmtable}. They agree excellently with results from other estimates from \textit{Gaia} data, but using a different method \citep[e.g.,][who also discussed the cluster internal dynamical properties]{kuhn2019}.

\begin{figure}
\includegraphics[width=0.49\textwidth]{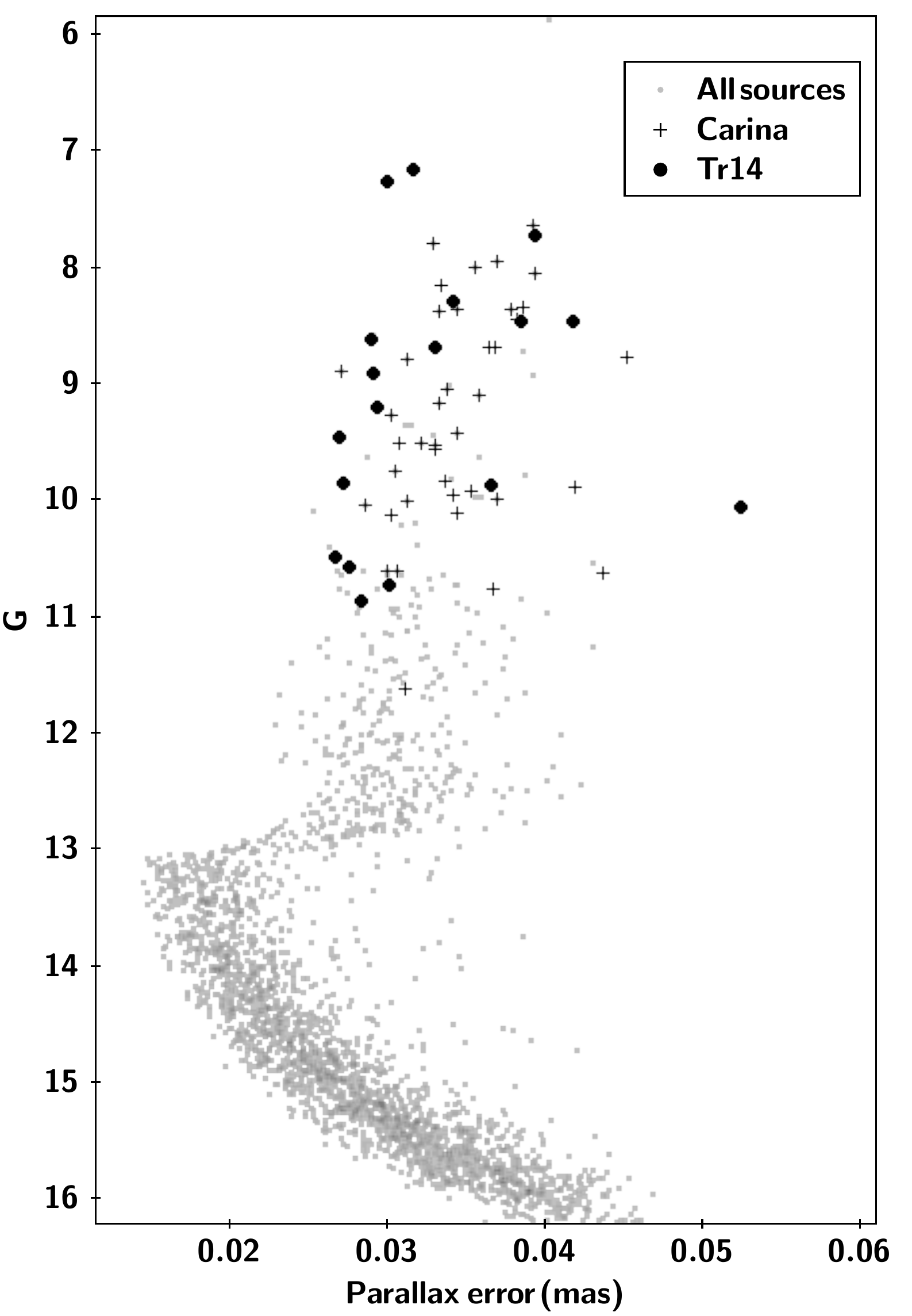}
\caption{ \textit{Gaia} G magnitude vs. parallax error. The gray points represent sources in the Trumpler-14, 15, and 16 region according to the selection criteria discussed in the text, while the black points (see inset for key) are the measurements for the bright OWN O-type stars in Carina and our Trumpler-14 candidate O-type stars.}
\label{parfigure}
\end{figure}

The mean parallaxes of the two clusters are identical to within $3\sigma$, having a value of 0.39 mas, to two significant figures.
Correcting for the systematic error in the \textit{Gaia} zero-point of 0.03 mas, but ignoring the uncertainty associated with the spatially correlated error term \citep[see discussion in subsection 5.4 of][]{Lindegren2018}, implies a mean parallax of 0.42$\pm0.05$ mas, or a distance of 2.38 kpc, with an uncertainty of about 10\%. This agrees very well with the $\eta$ Car geometric distance of 2.35$\pm$0.05 kpc \citep{smith2006}. 
In Fig.\ref{parfigure} we show the dependence of the parallax errors on magnitude, to be compared with Fig.\,B.2 of \citet{Lindegren2018}. The bright O-stars of interest here typically have parallax errors in the range 0.03 to 0.05 mas. 
Therefore distances to individual stars cannot be used to help define cluster membership, and we turn accordingly to the differences in proper motion between Trumpler-14 and Trumpler-16. 

\begin{table*}[t!]
\centering
\captionsetup{justification=centering}
\caption[center]{Mean parallax and proper motion estimates for Trumpler-14, 15, and 16, as discussed in Appendix A.}
\label{pmtable}
\centering
\begin{tabular}{lllllll}
\hline\hline
Cluster & \multicolumn{2}{c}{Field centers} & number & Parallax & pmRA & pmDEC \\
Name & RA & DEC & of stars & mas & mas\,yr$^{-1}$ & mas\,yr$^{-1}$\\ \hline
Trumpler-14 & 160.957 & $-59.568$ &   98  &  0.39 $\pm$0.03 & $-6.599\pm$0.227 &  1.980$\pm$0.156\\
Trumpler-15 & 161.167 & $-59.363$ & 92    &  0.39 $\pm$0.04 & $-6.185\pm$0.209 &  2.048$\pm$0.107\\
Trumpler-16 & 161.236 & $-59.720$ & 102   &  0.39 $\pm$0.04 & $-6.985\pm$0.153 &  2.591$\pm$0.108\\
\hline
\end{tabular}
\end{table*}

\begin{figure}
\includegraphics[width=0.49\textwidth]{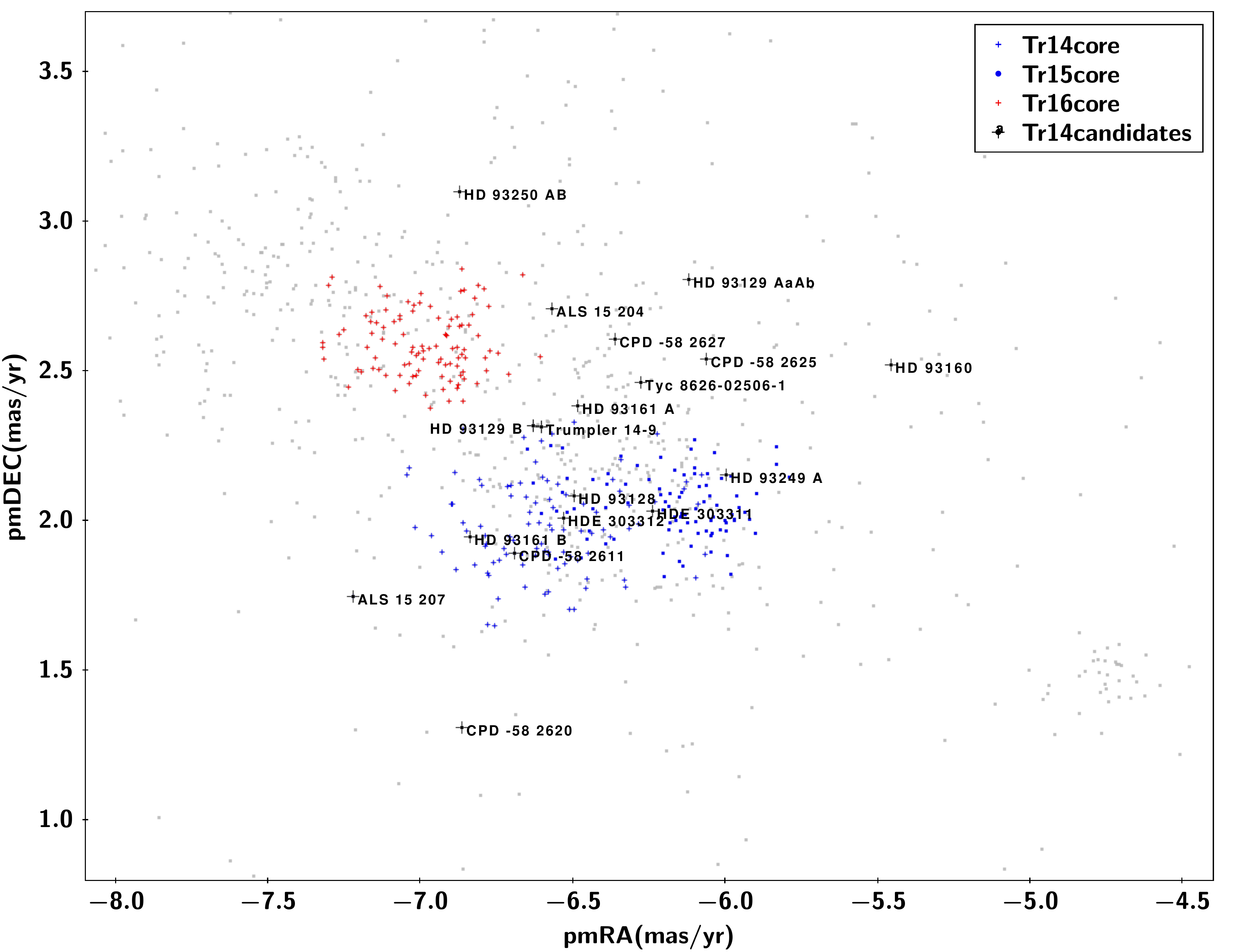}
\caption{Gray symbols are the proper motions of all sources that meet our filter criteria around the core Trumpler-14, 15, and 16 regions. The colored symbols represent the core dynamical groups as discussed in the text (see inset for key), after $2\sigma$ clipping. Overplotted are the proper motions of our candidate O-stars in Trumpler-14 (see Table \ref{Tr_14_Stars}). The two labels that overlap and are difficult to read are for the stars HD\,93129B and Trumpler-14-9. }
\label{pmfigure}
\end{figure}

We illustrate the proper motions in Figure \ref{pmfigure}, where we also show the proper motions of our candidate O-stars in Trumpler-14. The region clearly is dynamically quite complex, with significant substructure.
Nevertheless, based on the similarity of Trumpler-14 and Trumpler-15 mean proper motions, we can assign them to the same broad dynamical group, distinct from the Trumpler-16 group (which is clearly also part of a larger dynamical group extending to the upper left corner of Figure \ref{pmfigure}). We note here in passing that the anonymous group to the lower right in this figure consists of a comoving though spatially diffuse group of massive stars at the same distance as the other clusters.

While there are some significant outliers within the Trumpler-14 candidates to which we return below, the critical mid-O dwarf stars HDE\,303311, CPD-58\,2611, Tr 14-9, HD\,93161A, and HD\,93161B (cf. Table\ref{Tr_14_Stars}) are all securely within the Trumpler-14 and 15 group.
HD~93128 (an O3.5\,V((f))z) is a common Trumpler-14 member from the Gaia data, slightly more massive than the mid-O dwarfs crucial to the gap, and a member of the more massive stars that appear closer to the ZAMS.
The spectroscopic binary HD\,93160AB is rather discrepant, but it is in any case classified as O7\,III((f)). Its proper motion is even more inconsistent with Trumpler-16 membership, however (its proper motion vector is almost orthogonal to the Trumpler-16 proper motion relative to Trumpler-14). The proper motion of HD\,93219AaAb is also rather different from the two dynamical groups. In this case, however, we note that \citet{nelan2010} used HST/FGS to separate this system into two components separated by $53\pm3$ mas and $0.9\pm0.05$ magnitudes. Therefore it is quite possible that the closeness and comparable magnitudes of the components have affected the \textit{Gaia} astrometry and proper motion measurement \citep{lennon2018}. Finally, of the outliers we discuss here, we note that while HD\,93250 has also been separated into two components of almost equal magnitude \citep{sanahd93250}, with separation $1.5\pm0.2$ mas, its proper motion is similar to that of Trumpler-16, and it is likely a member of that cluster.

\section{Stellar tracks from stellar models in which  the accretion rate was altered}\label{AppHaemm}

In reference to the evolutionary models used in the Sect.~\ref{SectAccret}, we include here some of the characteristics that defines them. The Geneva stellar evolution code is a one-dimensional hydrostatic code that numerically solves the structure equations with the Henyey method.
The code includes energy production by gravitational contraction and nuclear reactions, including hydrogen-, lithium-, deuterium-, and all the hydrostatic burning phases until the core Si-burning phases.
Opacities are interpolated from the OPAL tables \citep{iglesias1996}, and convection is treated with the mixing-length theory and the Schwarzschild criterion.
A detailed description of the code without accretion is available in \cite{eggenberger2008}.

The treatment of accretion is described in detail in \cite{Haemmerle2016}.
Here we recall only the main ingredients.
The accretion rate is a free parameter.
The accreted material is added at each time step at the surface of the star.
Its thermal properties are set to be identical to that of the stellar surface (cold-disk accretion).
This assumption corresponds to a disk-like accretion geometry, in which any entropy excess is radiated away in the polar directions
before it is advected in the stellar interior.
This is a lower limit for the accretion of entropy that leads during the early accretion phase to smaller stellar radii than any other assumption.
However, the differences in radii between hot and cold accretion disappear as the star reaches the ZAMS (ignition point),
so that we do not expect this assumption to affect the results of our study.

The main characteristics of the models used are included in Table~\ref{GenMod}.
We first consider constant accretion rates of $\dot M=10^{-5}-10^{-4}-10^{-3}\rm\,M_\odot\,yr^{-1}$.
For $\dot M=10^{-5}-10^{-4}\rm\,M_\odot\,yr^{-1}$, the runs start at $0.7\rm\,M_\odot$.
However, accretion at high rate on low-mass objects makes numerical convergence difficult to achieve.
For $\dot M=10^{-3}\rm\,M_\odot\,yr^{-1}$ we therefore start the computation at $2\rm\,M_\odot$.
In all the cases, the initial protostellar seed is a fully convective hydrostatic object located at the top of the Hayashi line corresponding to its mass.
All the accreting models run until they reach a final mass of $M=70\rm\,M_\odot$.
For the chemical composition, we use solar abundances (Z=0.014, \citealt{asplund2005,cunha2006}) in the initial model and the accreted material.
In order to distinguish between the observational constraints on the various mass ranges,
we compute in addition models with initial accretion rates identical to those described above,
but switched to a different value when the stellar mass reaches $25\rm\,M_\odot$. 
A detailed description of the internal and surface properties of the models at constant rates is available in \cite{Haemmerle2016}.

%=============================================================================================================
\begin{table}[!t]
        \caption{Geneva evolutionary models with controlled mass-accretion rate used here} 
        \label{GenMod}
        \centering
            \begin{tabular}{rccl} %lr@{\hskip 0.05in}lccccc
                \hline \hline
        \noalign{\smallskip}
                Name &   $\dot{M}_{init}$ & $\dot{M}_{>25M_{\odot}}$       & Draw  \\
                     &   [M$_{\odot}$ yr$^{-1}$] & [M$_{\odot}$ yr$^{-1}$] &       \\
                \hline
        \noalign{\smallskip}
    Model1 & 10$^{-5}$   & 10$^{-5}$  & Fig. 9 (green)  \\
    Model2 & 10$^{-5}$   & 10$^{-4}$  & Fig. 9 (red)    \\
    Model3 & 10$^{-4}$   & 10$^{-4}$  & Fig. 10 (red)    \\
    Model4 & 10$^{-4}$   & 10$^{-5}$  & Fig. 10 (green)  \\
    Model5$^{**}$ & 10$^{-3}$   & 10$^{-3}$  & Fig. 11 (blue)   \\  %\tnote{a}
    Model6$^{**}$ & 10$^{-3}$   & 10$^{-5}$  & Fig. 11 (green)  \\
        \hline
        \noalign{\smallskip}
        \multicolumn{4}{l}{Initial mass: 0.7 \msol. ($^{**}$ 2 \msol). Final mass: 70 \msol.} \\
        \multicolumn{4}{l}{Z=0.014} \\
        \end{tabular}
\end{table}
%=============================================================================================================

\section{Proxy shape of the gap}\label{AppRedline}

As a quick method to favor comparisons with our resulting gap near the ZAMS, we designed a polyhedron using specific points of the evolutionary tracks in the sHRD (Fig.~\ref{VGAP}) surrounding the region. Table~\ref{RedLine} includes the parameters that define the models at these points for the sHRD. We then transferred the shape to the remaining diagrams (Kiel and HR) with the remaining parameter values for that model at the same point, also included in Table~\ref{RedLine}.

%=============================================================================================================
\begin{table}[!t]
        \caption{Parameter model values to define a proxy of the shape of the gap near the ZAMS} 
        \label{RedLine}
        \centering
            \begin{tabular}{ccccc} 
                \hline \hline
        \noalign{\smallskip}
                Mass at ZAMS &   log($T_{\rm eff}$) & \grav       &  log($\mathcal{L}$/$\mathcal{L}_\odot$)  & log(L/L$_\odot$)  \\
                  \phantom{g} [\msol] &         &       &             &                                              \\
                \hline
        \noalign{\smallskip}
    25  &   4.59  &  4.28   &    3.47   &    4.86   \\ 
    30  &   4.60  &  4.10   &    3.69   &    5.19   \\
    60  &   4.65  &  3.97   &    4.01   &    5.76   \\
    85  &   4.71  &  4.19   &    4.05   &    5.98   \\
        \hline
        \noalign{\smallskip}
        \end{tabular}
\end{table}
%=============================================================================================================

\section{Tables}

%%285 Stars, check. 
%
%
% -----------------------------------------------------------------------
% --------------------           TABLE A .1       ------------------------
% ----------------------------------------------------------------------- 
%        \setlength{\footskip}{60pt}
%        \begin{landscape}
%        \pagestyle{empty}
%\fontsize{8}{8}\selectfont

\onecolumn

%\begin{multicols}{2}
%\raggedleft
{
\setlength\LTleft{-0.5cm}  %For longauth

\noindent
\begin{longtable}{lr@{\hskip 0.05in}l@{\hskip 0.1in}cccccl}
\caption{List of the 285 stars with spectroscopic parameters obtained in this work. Stars are separated by luminosity class (horizontal lines) and sorted by spectral type. Columns are: (1, 2) Name and spectral type from GOSC, (3, 4, 5) \Teff, \gravt\ and log~$\mathcal{L}$ from this work \citep[See notes on][for associated uncertainties]{Holgado2018}, (6) $B$ magnitude from GOSC, (7) the extinction measurement $E$(4405-5495) from \cite{MaizApellaniz2018}, (8) additional notes.}\\
\label{tableValues}\\
\hline \hline
\noalign{\smallskip}
Name &   SpT &             LC &      $T_{\rm eff}$ & \gravt & log $\mathcal{L}$/$\mathcal{L}$$_{\odot}$ & $B$ & E(4405-5495) & Notes \\
     &       &                &           [kK]     & [dex] &            [dex]                          & [mag]  &           &   \\
\hline
\noalign{\smallskip}
\endfirsthead
\caption{continued.}\\
\hline
\hline
\noalign{\smallskip}
Name &   SpT &             LC &      $T_{\rm eff}$ & \gravt &  log $\mathcal{L}$/$\mathcal{L}$$_{\odot}$ & $B$ & E(4405-5495) & Notes \\
     &       &                &           [kK]     & [dex] &             [dex]                          & [mag]  &           &       \\
\hline
\noalign{\smallskip}
\endhead
%\hline
%\hline \hline
\noalign{\smallskip}
\noalign{\smallskip}
\hline\hline
\noalign{\smallskip}
\multicolumn{9}{l}{$\mathcal{L}=T_{\rm eff}^4$\,/$g_{\rm true}$,\, $g_{\rm true}=g+g_{\rm cent}$, $g_{\rm cent}\approx\frac{(V_{\rm rot}\sin i)^2}{R_{*}} $}\\
\endfoot
    HD\,93129\,AaAb           &    O2 &          I\,f* &    45.6$\pm$1.1 &     3.89$\pm$0.10 &      4.14$\pm$0.11  &       7.8 &        0.514      &  Weak \ioni{He}{i} lines; SB1     \\
         CYG\,OB2-7           &    O3 &          I\,f* &    50.3$\pm$1.8 &     4.09$\pm$0.13 &      4.11$\pm$0.14  &      12.2 &        1.827      &  Weak \ioni{He}{i} lines    \\
         ALS\,15210           &  O3.5 &     I\,f*\,Nwk &    42.4$\pm$1.6 &      3.69$\pm$0.05 &     4.21$\pm$0.08  &      11.5 &            .      &  Weak \ioni{He}{i} lines    \\
      HD\,190429\,A           &    O4 &           I\,f &    36.6$\pm$0.8 &     3.53$\pm$0.09 &      4.11$\pm$0.10  &       7.3 &        0.447      &  .    \\
          HD\,15570           &    O4 &           I\,f &    40.2$\pm$0.8 &     3.63$\pm$0.06 &      4.18$\pm$0.07  &       8.8 &        0.933      &  Weak \ioni{He}{i} lines    \\
          HD\,16691           &    O4 &           I\,f &           $>$39.7 &         $>$3.74 &            $<$4.05  &       9.1 &        0.742      &  Weak \ioni{He}{i} lines    \\
          HD\,14947           &  O4.5 &           I\,f &    39.1$\pm$1.1 &     3.65$\pm$0.10 &      4.11$\pm$0.11  &       8.5 &        0.689      &  .    \\
         CYG\,OB2-9           &  O4.5 &           I\,f &    40.1$\pm$1.0 &     3.92$\pm$0.10 &      3.88$\pm$0.11  &      12.9 &        2.293      &  Weak \ioni{He}{i} lines    \\
\end{longtable}
   }

%\onecolumn
%{
%%\setlength\LTright{+5cm}  %For nothing
%\noindent\begin{longtable}{lr@{\hskip 0.05in}l@{\hskip 0.1in}cccccl}
%\label{tableValues}\\
%   \end{longtable}
%}
%

\onecolumn

% -----------------------------------------------------------------------
% --------------------           TABLE A .2       ------------------------
% ----------------------------------------------------------------------- 
%        \setlength{\footskip}{60pt}
%        \begin{landscape}
%        \pagestyle{empty}
%\fontsize{8}{8}\selectfont
%\begin{multicols}{2}
%\raggedleft
{
\noindent\begin{longtable}{lr@{\hskip 0.05in}lccc}
\caption{113 O-type stars in the sample with double-line spectroscopic features, separated by luminosity class (I+II, III, and IV+V) and sorted by spectral type. No analysis available. It details name, approximate spectral classification \citep[sometimes is known to be a compromise, as the case of HD\,37468, an SB3 star][]{Simon-Diaz2015a,MaizApellaniz2018a}, $B$ magnitude, reddening parameter $E$(4405-5495), and secondary component spectral classification in the unresolved spectrum. Everything listed in GOSC \citep{MaizApellaniz2013} and \cite{MaizApellaniz2018}.}\\
\label{tableValues_NoAnalisis}\\
\hline \hline
\noalign{\smallskip}
Name &   SpT &             LC &   $B$ & E(4405-5495) & LC  \\
     &       &                &   [mag]    &              & comp. \\ 
\hline
\noalign{\smallskip}
\endfirsthead
\caption{continued.}\\
\hline
\hline
\noalign{\smallskip}
Name &   SpT &             LC &   $B$ & E(4405-5495) & LC  \\
     &       &                &   [mag]    &              & comp. \\ 
\hline
\noalign{\smallskip}
\endhead
%\hline
%\hline \hline
\noalign{\smallskip}
\noalign{\smallskip}
\hline\hline
\noalign{\smallskip}
\endfoot
            HD\,93162 &      O2.5 &           I\,f*/WN6  &   8.5 &   0.695 &  OB         \\   
       LS\,III\,+4611 &      O3.5 &               I\,f*  &  12.5 &   1.726 &  O3.5If*    \\   
         HD\,228\,766 &        O4 &               I\,f*  &   9.8 &   0.884 &  O8:II:     \\   
          HDE\,229196 &        O6 &             II\,(f)  &   9.4 &   1.158 &  .          \\   
           HD\,153919 &        O6 &             Ia\,fcp  &   6.8 &   0.499 &  .          \\   
           HD\,150958 &      O6.5 &            Ia\,(n)f  &   7.9 &   0.613 &  .          \\   
           HD\,152248 &        O7 &              Iab\,f  &   6.3 &   0.403 &  O7Ib(f)    \\   
           HD\,101205 &        O7 &            II:\,(n)  &   6.6 &   0.322 &  .          \\
\end{longtable}
}

% -----------------------------------------------------------------------
% --------------------           TABLE A .2       ------------------------
% ----------------------------------------------------------------------- 
%        \setlength{\footskip}{60pt}
%        \begin{landscape}
%        \pagestyle{empty}
%\fontsize{8}{8}\selectfont
\onecolumn
%\begin{multicols}{2}
%\raggedleft
{
\noindent\begin{longtable}{lr@{\hskip 0.05in}lccr}
\caption{17 peculiar stars (Oe, WR, and Mag) in the sample with no spectroscopic analysis available, separated by luminosity class (I+II, III, and IV+V) and sorted by spectral type. Columns include: name, spectral class, $B$ magnitude, reddening parameter $E$(4405-5495), and the reason that makes the analysis of the star impossible, i.e., the star is Oe, Magnetic or WR. Magnetic stars include reference to its magnetic detection.}\\
\label{tableValues_NoAnalisis_MagWR}\\
\hline \hline
\noalign{\smallskip}
Name &   SpT &             LC  &   $B$ & E(4405-5495) & Notes \\
     &       &                 &   [mag]    &              & Notes \\ 
\hline
\noalign{\smallskip}
\endfirsthead
\caption{continued.}\\
\hline
\hline
\noalign{\smallskip}
Name &   SpT &             LC  &   $B$ & E(4405-5495)  & Notes \\
     &       &                 &   [mag]    &               &  \\ 
\hline
\noalign{\smallskip}
\endhead
%\hline
%\hline \hline
\noalign{\smallskip}
\noalign{\smallskip}
\hline\hline
\noalign{\smallskip}
\multicolumn{4}{l}{ Notes: Mag:Magnetic, WR: Wolf-Rayet.} \\
\multicolumn{5}{l}{*: \cite{Grunhut2017}, **: \cite{Caballero-Nieves2009}.}\\
\multicolumn{5}{l}{***: \cite{Hubrig2008},****: \cite{Hubrig2011}, º: \cite{Martins2010}.}\\
\multicolumn{5}{l}{º*: \cite{Wade2011}, º**: \cite{Naze2008}.} \\

\endfoot
            HD\,66811 &        O4 &            I\,(n)fp &    2   &   0.012 &  Mag*   \\
            ALS\,2063 &        O5 &               I\,fp &   11.5 &   1.027 &  WR   \\
           HD\,152386 &       O6: &             Ia\,fpe &    8.6 &   0.775 &  WR   \\
           HD\,313846 &       O7: &             Ia\,fpe &   10.7 &   1.113 &  WR   \\
           HD\,152408 &       O8: &             Ia\,fpe &    5.9 &   0.414 &  WR   \\
           HD\,226868 &      O9.7 &          Iab\,p\,var &   9.7 &   1.042 &  Mag**  \\
\hline
\noalign{\smallskip}
           HD\,148937 &        O6 &            ...\,f?p &    7.8 &   0.619 &  Mag***   \\
       CPD\,-28\,2561 &      O6.5 &              .\,f?p &   10.1 &   0.439 &  Mag****   \\
\end{longtable}
}

	\end{document}